\documentclass[%
 aip,
 amsmath,amssymb,
 reprint,%
]{revtex4-1}

\usepackage{graphicx}
\usepackage{dcolumn}
\usepackage{bm}

\usepackage[utf8]{inputenc}
\usepackage[T1]{fontenc}
\usepackage{mathptmx}
\usepackage{etoolbox}

\makeatletter
\def\@email#1#2{%
 \endgroup
 \patchcmd{\titleblock@produce}
  {\frontmatter@RRAPformat}
  {\frontmatter@RRAPformat{\produce@RRAP{*#1\href{mailto:#2}{#2}}}\frontmatter@RRAPformat}
  {}{}
}%
\makeatother
\begin{document}

\preprint{AIP/123-QED}
\title[]{Transient growth in a flat plate boundary layer under a stream with uniform shear}
\author{Shyam Sunder Gopalakrishnan}
\affiliation{ 
Laboratoire de Physique des Lasers, Atomes et Mol\'{e}cules, CNRS UMR 8523, Universit\'{e} Lille 1 - 59655 Villeneuve d'Ascq Cedex, France
}%
\affiliation{ 
Facult\'{e} des Sciences, Universit\'{e} libre de Bruxelles (ULB), CP. 231, 1050 Brussels, Belgium
}%
\author{Alakesh Chandra Mandal}%
 \email{shyam7sunder@gmail.com, alakeshm@iitk.ac.in}
\affiliation{ 
Department of Aerospace Engineering, Indian Institute of Technology, Kanpur 208016, India
}%

\date{\today}

\begin{abstract}
One of the simplest problems involving external vorticity in boundary layer flows is the flow over a semi-infinite plate under a stream of uniform shear. We study the transient growth phenomenon in this flow to investigate the role of freestream shear on energy amplification, and analyse the differences with the Blasius flow. The initial optimal disturbance which triggers the maximum growth is found to be streamwise vortices, as in other shear flows. Compared to the Blasius boundary layer, higher optimum energy and larger spanwise wavelength of streamwise vortices have been observed. We provide scaling laws for the maximum optimal amplification, which is found to increase exponentially with the freestream shear gradient.
\end{abstract}

\maketitle

\section{\label{sec:Intro} Introduction}

The effects of freestream vorticity and/or turbulence on a laminar boundary layer has been of academic interest for many a decade as it finds its relevance in diverse engineering applications \cite{klebanoff1971effect,kendall1998experiments,westin1998experiments,saric2002boundary,schrader2010receptivity, Manu2010}. In flows which are subject to a high freestream disturbance environment, transition processes may bypass the linear mechanism altogether, and can involve nonlinearity directly \citep{Mork69}. It is by now well-established that the problem of receptivity to freestream turbulence is directly coupled to the transient growth of non-orthogonal modes \cite{klebanoff1971effect,jonavs2000receptivity,saric2002boundary,Henningson2006}. Here we study the transient growth in a flat plate boundary layer subject to a uniform freestream shear, which is one of the simplest problems involving external vorticity \cite{van1969higher}.

Under the linear approximation, Kovasznay \cite{Kova53} has shown that a small amplitude unsteady disturbance in the freestream can be decomposed into an acoustic, vortical, and entropy components. The pressure fluctuation associated with any acoustic disturbance can cause instability. On the other hand, vortical disturbances do not create a pressure gradient at the boundary layer edge and so do not provoke flow instability. However, vortical  disturbances can interact with surface irregularities leading to instability \cite{Crou94}. Freestream turbulence and wake in the freestream often contains a large amount of vortical disturbances, and undergo bypass transition \cite{Ovch06,Pan08}. 
A simple flow with freestream vorticity is the flow past a semi-infinite plate in a stream with uniform shear, 
which arose as a practical problem in supersonic flow past a blunt body \cite{ferri1954note,van1969higher}. 
Several investigators have studied the role of freestream shear on the boundary layer \cite{Li56,Glau57,Murr61,devan1965approximate,koch1971diffusion,Dey84,Bera05,legner2014optimal,balamurugan2017experiments1}. 
The existence of a streamwise pressure gradient due to interaction between the displacement thickness and external vorticity was first pointed out by Li \cite{Li56}. This was later confirmed using the method of matched asymptotic expansions by Murray \cite{Murr61} with the interaction between the displacement thickness and external vorticity stemming from the second-order boundary layer effect. It was further extended by Toomre \& Rott \cite{toomre1964pressure} wherein it was shown that the interaction pressure gradients were correctly given only while the assumption of an infinite, uniform shear flow is valid, and is strongly influenced by the boundedness of the external shear. A detailed review on this flow along with other relevant references can be found in this excellent article \cite{van1969higher}.

In the aforementioned boundary layer flows subject to freestream vortical disturbances, transition scenarios occur on a short time scale, and may bypass the linear mechanism \cite{tumin2001spatial,fransson2004experimental,mans2007sinuous,duriez2009self,balamurugan2017experiments2,vavaliaris2020optimal,rigas2021nonlinear}. The modal stability theory  which describes the asymptotic fate of infinitesimal disturbances \cite{drazin2004hydrodynamic}, fails to capture the short-term characteristics. On the other hand, non-modal stability theory describes the stability over a finite-time horizon \cite{schmid2007nonmodal}. The transient growth of energy is also referred to as non-modal since it is caused due to the superposition of several eigenmodes. Mathematically, the transient growth is due to the non-normal nature of the stability operator \cite{Farr88,farrell1988optimal,Redd93,Butl92}, which results in non-orthogonal eigenfunctions leading to an algebraic growth for a short time. This growth may be sufficient enough to trigger nonlinearity in flow and hence, transition at a subcritical Reynolds number \cite{Schm00,schmid2007nonmodal}. Transient growth phenomenon has been extensively studied for many shear flows like the Blasius boundary layer \cite{Elli75,Land80,Hult81,Ande99,Gust91,Butl92,Corb00,luchini2000reynolds,fransson2004experimental,hoepffner2005transient,zuccher2006algebraic,
 vavaliaris2020optimal}.
 
The linear stability of a boundary layer flow over a flat plate subject to a uniform freestream shear was studied by Bera \& Dey\cite{Bera05}. They showed that the freestream shear results in larger stability domains in comparison with the Blasius boundary layer flow. However, the non-normal nature of the governing operator may result in significant transient growth, which forms the objective of this study. Recently, the effect of mean flow shear in flat plate boundary layers was experimentally investigated in a low-speed wind tunnel \cite{balamurugan2017experiments1}. By using a non-uniform parallel rod grid, they were able to control the freestream turbulence resulting in a mean velocity profile with shear. They showed that the energy amplification is higher compared to mean velocity profiles without shear, thereby undergoing bypass transition earlier. In addition, they further observed algebraic growth of distrubance energy within the boundary layer. In the present work we perform a transient growth analysis of such a flow with uniform freestream shear, which has a second-order boundary layer effect. As observed in \cite{balamurugan2017experiments1}, we find that the transient amplification of the disturbances scale with the imposed freestream shear.
 
To this end, we organise the manuscript as follows. The governing equations along with the mean flow velocity profiles are discussed in \ref{sec:goveqn}. The tools used in non-modal stability analyses are also briefly outlined for completeness. Following this, we present the results outlining the differences in energy amplification in flows with and without shear (\ref{sec:res}). We also provide scaling laws for the transient growth as a function of the imposed freestream shear, which is followed by some concluding remarks in \ref{sec:conc}.

\section{Problem formulation and methodology}\label{sec:goveqn}

In the following, $x$, $y$ and $z$ denote the streamwise, wall-normal and spanwise directions, respectively. 
A simple model representative of freestream vorticity is a steady incompressible laminar flow over a semi-infinite plate placed in a stream with uniform shear, with the freestream velocity given by $U_{s}=U_{0}+\kappa y$. Here $U_{0}$ and $\kappa$ are constants, with the uniform freestream Blasius flow corresponding to $\kappa = 0$. Since the freestream vorticity introduces an additional second-order boundary layer effect, this flow has been studied by several authors \cite{Li56,Glau57,Murr61,koch1971diffusion,Dey84}.
 For the flow under consideration, the imposed freestream shear is $dU_{s}/dy$, the non-dimensional vorticity number being $N_{1}(=\kappa \nu/U_{0}^{2})\ll$ 1. The stream function, $\psi$, may be expressed as the perturbation about the Blasius flow,
\begin{equation}
\psi(x,y)=\sqrt{U_{0}x\nu}\big[f_{0}(\eta)+N_{1}\sqrt{(Re)}f_{1}(\eta)+....\big]
\end{equation}
where $\eta =y\sqrt{U_{0}/\nu x}$, $f_{0}(\eta)$ is the non-dimensional Blasius stream function, and $f_{1}(\eta)$ is the non-dimensional perturbed stream function. The governing similarity equations for $f_{0}(\eta)$ and $f_{1}(\eta)$ that follows from the boundary layer momentum equations \citep{Murr61} are
\begin{equation}
f^{\prime\prime\prime}_{0}+f_{0}f^{\prime\prime}_{0}=0,
\end{equation}

\begin{equation}
f^{\prime\prime\prime}_{1}+f_{0}f^{\prime\prime}_{1}-f^{\prime}_{0}f^{\prime}_{1}+2f^{\prime\prime}_{0}f_{1}=A.
\end{equation}

Here a prime denotes the derivative with respect to $\eta$, and the constant $A$ is due to the imposed pressure. The boundary condition for the above set of equations are as follows.
\begin{equation}
f_{0}(0)=f^{\prime}_{0}(0)=0,\qquad f^{\prime}_{0}(\infty)\rightarrow 1 
\end{equation}
\begin{equation}
f_{1}(0)=f^{\prime}_{1}(0)=0,\qquad f^{\prime}_{1}(\infty)\rightarrow \eta 
\end{equation}

  \begin{figure}
 \unitlength=50.0mm
\centerline{
\includegraphics[width=1.58\unitlength,height=1.18\unitlength]{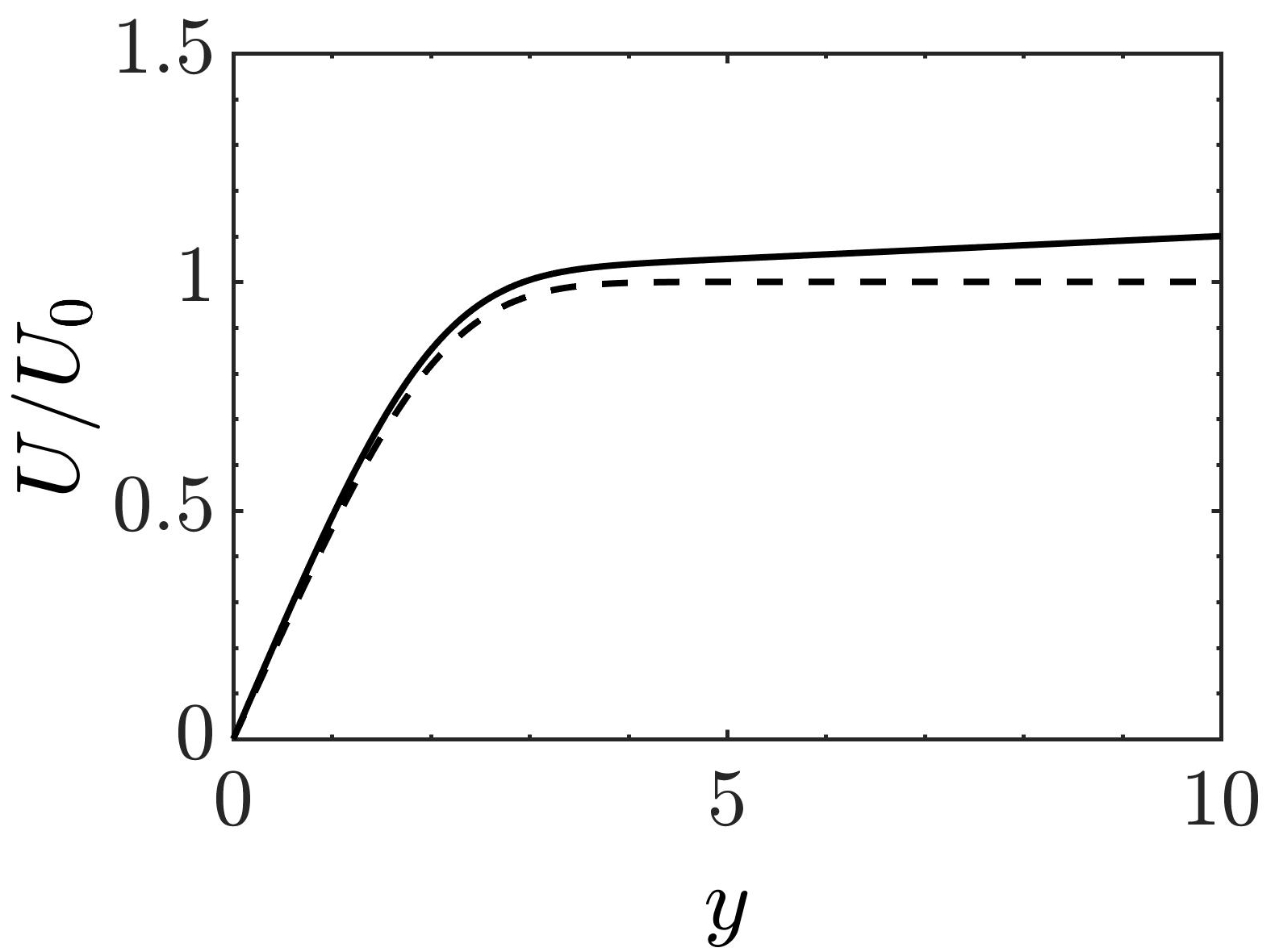}
}
 \caption{Boundary layer velocity profiles. - - - - Blasius; ----- uniform shear, for $N$ = 0.01.
}
 \label{fig:velo}
 \end{figure}  		

The base flow used in the transient growth analysis is given by $U = f^{\prime}(\eta)=f^{\prime}_{0}(\eta)+Nf^{\prime}_{1}(\eta)$, where $N=N_{1}Re/1.73$. Figure~\ref{fig:velo} shows the mean velocity profiles with and without the uniform freestream shear. It can be noted  that the principal effect of the external vorticity is to result in an increasing displacement thickness \cite{Murr61}. This results in a modified pressure field outside the boundary layer region, thereby affecting the skin friction, boundary layer separation, and stability region. Indeed, the linear stability analysis has shown that the freestream shear stabilizes the flow in comparison with the Blasius flow without any shear \cite{Bera05}.

We  now briefly outline the tools used in the non-modal stability analysis for completeness. For the transient growth analysis on the above base flow, which we denote as $\bf{U}$ = $(U, 0, 0)$, we add a three-dimensional disturbance with the velocities given by $\mathbf {\tilde{u}} = (\tilde{u}, \tilde{v}, \tilde{w})$. The velocities are normalized by $U_{0}$, and distances by the displacement thickness of the Blasius boundary layer. Pressure has been normalized by $\rho U_{0}^{2}$. Under the parallel flow assumption, the linearized disturbance equations and the continuity equation reads as
\begin{equation}
\frac{\partial \tilde{u}}{\partial t}+U\frac{\partial \tilde{u}}{\partial x}+vU'=-\frac{\partial p}{\partial x}+\frac{1}{Re}\nabla^{2}\tilde{u},
\end{equation}
\begin{equation}
\frac{\partial \tilde{v}}{\partial t}+U\frac{\partial \tilde{v}}{\partial x}=-\frac{\partial p}{\partial y}+\frac{1}{Re}\nabla^{2}\tilde{v},
\end{equation}
\begin{equation}
\frac{\partial \tilde{w}}{\partial t}+U\frac{\partial \tilde{w}}{\partial x}=-\frac{\partial p}{\partial z}+\frac{1}{Re}\nabla^{2}\tilde{w},
\end{equation}

\begin{equation}
\frac{\partial \tilde{u}}{\partial x}+\frac{\partial \tilde{v}}{\partial y}+\frac{\partial \tilde{w}}{\partial z}=0.
\end{equation}

Considering the normal vorticity
\begin{displaymath}
\tilde{\Omega}=\frac{\partial \tilde{u}}{\partial z}-\frac{\partial \tilde{w}}{\partial x},
\end{displaymath}
 and eliminating the perturbation pressure, we obtain the following two equations \citep{Schm01}
\begin{equation}
\Big[\Big(\frac{\partial}{\partial t}+U\frac{\partial}{\partial x}\Big)\nabla^{2}-U^{\prime\prime}\frac{\partial}{\partial x}-\frac{1}{Re}\nabla^{4}\Big]\tilde{v}=0,
\end{equation}
\begin{equation}
\Big[\frac{\partial}{\partial t}+U\frac{\partial}{\partial x}-\frac{1}{Re}\nabla^{2}\Big]\tilde{\Omega}=-U^{\prime}\frac{\partial\tilde{v}}{\partial z}.
\end{equation}

The boundary conditions are $\tilde{v}=\tilde{v^{\prime}}=\tilde{\Omega}=0$. The solutions of these equations can be sought as follows \citep{Schm01}
\begin{equation}
\tilde{v}(x,y,z,t)=v(y,t)e^{i(\alpha x+\beta z)},
\end{equation} 
\begin{equation}
\tilde{\Omega}(x,y,z,t)=\Omega(y,t)e^{i(\alpha x+\beta z)}.
\end{equation}

Here we consider the homogeneous nature of the flow in the streamwise and the spanwise coordinate directions, with $\alpha$ and $\beta$ denoting the corresponding wave numbers. By substituting these velocity and vorticity in (10) and (11), 
we obtain 
\begin{equation}
\Big[\big(\frac{\partial}{\partial t}+i\alpha U\big)\big(D^{2}-k^{2}\big)-i\alpha U^{\prime\prime}-\frac{1}{Re}\big(D^{2}-k^{2}\big)^{2}\Big]v=0,
\end{equation}
\begin{equation}
\Big[\big(\frac{\partial}{\partial t}+i\alpha U\big)-\frac{1}{Re}\big(D^{2}-k^{2}\big)\Big]\Omega=i\beta U^{\prime}v.
\end{equation}
along with the boundary conditions $v=Dv=\Omega=0$ at the wall and in the far field.
Here, $D=\frac{d}{dy}$ and $k=\sqrt{\alpha^{2}+\beta^{2}}$.

Using the continuity equation, the horizontal velocity components, $u$ and $w$, can be recovered 
from the normal velocity and the normal vorticity by the following equations.
\begin{equation}
u=\frac{i}{k^{2}}\big(\alpha Dv-\beta \Omega\big),
\end{equation} 
\begin{equation}
w=\frac{i}{k^{2}}\big(\beta Dv+\alpha \Omega\big).
\end{equation}

The equations (14) and (15) can also be written in vector notation \citep{Schm01}
\begin{equation}
\mathbf M \frac{\partial\mathbf q}{\partial
t}=\mathbf{L}\mathbf{q} \qquad \mathrm{or} \qquad
\frac{\partial\mathbf q}{\partial t}=\mathbf
M^{-1}\mathbf{L}\mathbf{q}=\mathbf{L_{1}\mathbf{q}},
\end{equation}
where the vectors $\bf q$, $\bf M$, and $\bf L$ are \\
\\
 $\bf{q}=\big(\begin{array}{c}v \\ \Omega \end{array}\big)$, \qquad
$\bf{M}=\left(\begin{array}{cc} k^{2}-D^{2} & 0 \\ 0 & 1 \end{array} \right)$ \quad and \quad \\
 $\bf{L}=\left(\begin{array}{cc} L_{OS} & 0 \\ i\beta U^{\prime} & L_{SQ} \end{array} \right)$.\\

The Orr--Sommerfeld operator, $L_{OS}$, and the Squire operator, $L_{SQ}$, are
\begin{displaymath}
L_{OS}=i\alpha U\big(D^{2}-k^{2}\big)-i\alpha U^{\prime\prime}-\frac{1}{Re}\big(D^{2}-k^{2}\big)^{2},
\end{displaymath}
\begin{displaymath}
L_{SQ}=i\alpha U^{\prime\prime}-\frac{1}{Re}\big(D^{2}-k^{2}\big).
\end{displaymath}
Assuming solutions of the form
\begin{displaymath}
\mathbf{q} = \mathbf{\tilde{q}} \exp(-i \omega t)  \qquad
\qquad \mathbf{\omega \in C}
\end{displaymath}
the initial value problem (eq. 13) becomes a
generalized eigenvalue problem 
\begin{equation}
\mathbf{L}\mathbf {\tilde{q}} = -i \omega \mathbf{M}
\mathbf{\tilde{q}}.
\end{equation}
The energy measure of the disturbance is
\begin{equation}
E(t)=\frac{1}{2k^{2}}\int_{0}^{\infty} \big(\begin{array}{c}v \\ \Omega \end{array}\big)^{H}\Big(\begin{array}{cc} k^{2}-D^{2} & 0 \\ 0 & 1 \end{array} \Big)\big(\begin{array}{c}v \\ \Omega \end{array}\big) \ dy,
\end{equation} 
and the maximum possible amplification of the initial energy density \citep{Schm01} is
\begin{equation}
G(t)=\max_{\bf q_{0}\neq 0}\frac{\bf \Vert q(t)\Vert^{2}}{\bf \Vert q_{0}\Vert^{2}}
\end{equation} 
where $\Vert \bf{q(t)}\Vert^{2}$ = $E(t)$.

Here $G(t)$ is the maximum possible energy amplification at a given time, and includes an optimisation over all initial conditions. At different time instants a different initial condition may yield the maximum possible energy amplification. The interested reader is referred to these excellent references \cite{Schm01,schmid2007nonmodal} for further details.

\section{Results and Discussion}\label{sec:res}

  \begin{figure}
 \unitlength=50.0mm
\centerline{
\includegraphics[width=1.55\unitlength,height=1.17\unitlength]{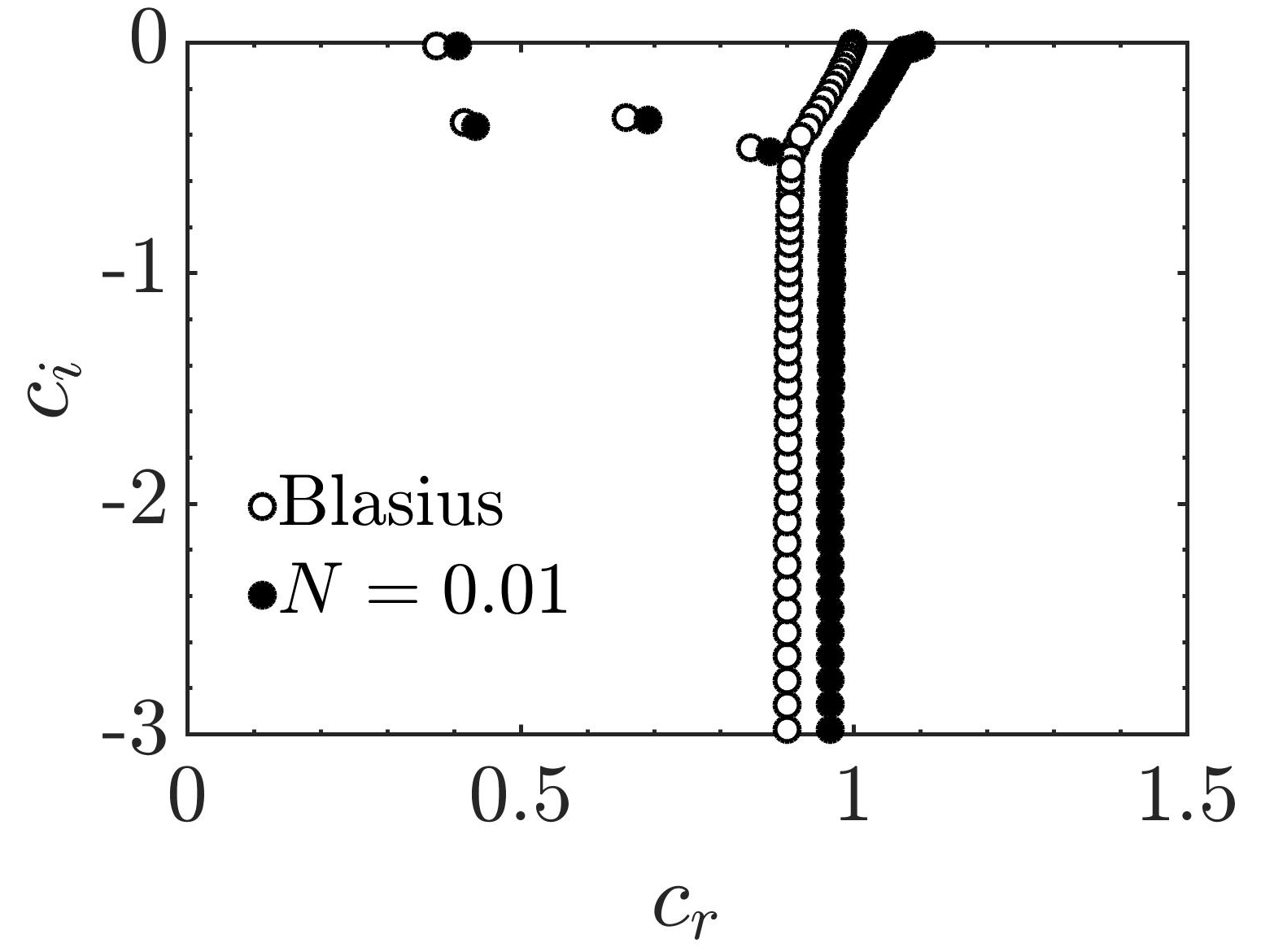}
}
 \caption{
Eigenvalue spectra for Blasius and uniform freestream shear flows at $\alpha = 0.2$ and $Re=500$.
}
 \label{fig:eig}
 \end{figure} 
 
   \begin{figure}
 \unitlength=45.0mm
\centerline{
\includegraphics[width=1.64\unitlength,height=1.17\unitlength]{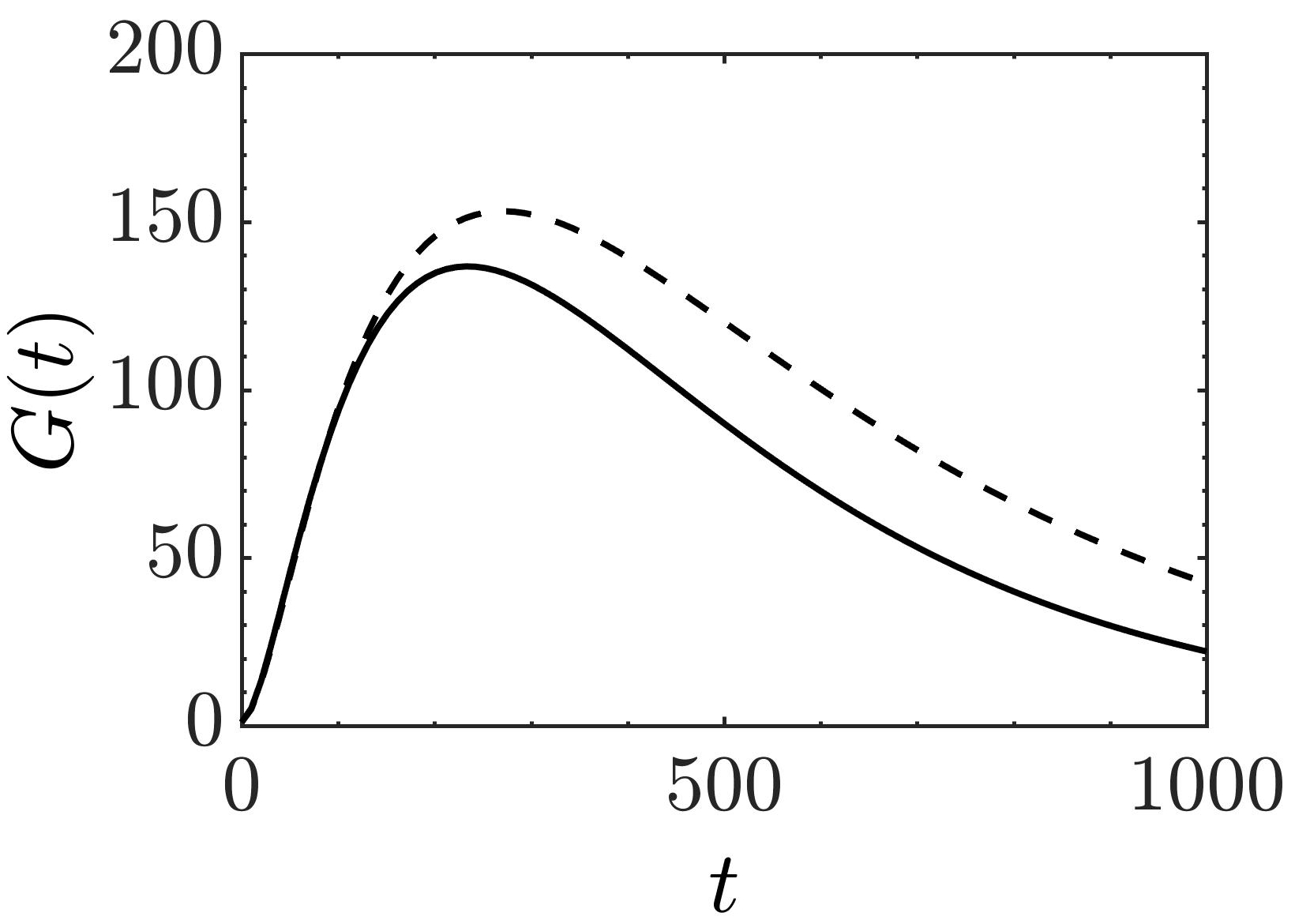}
}
 \caption{
Maximum amplification, $G(t)$, for Blasius ($N$ = 0; solid line) and uniform freestream shear ($N = 0.01$; dashed line) flows at $Re = 300$. $\alpha = 0$ and $\beta$ $\approx$ 0.65 for $N = 0$; $\alpha = 0$ and $\beta$ $\approx$ 0.60 for $N = 0.01$.
}
 \label{fig:gt}
 \end{figure}

Using a spectral code, the eigenfunctions and eigenvalues of the stability operator, $\bf L_{1}$, are obtained. An arbitrary initial disturbance may be obtained by using the eigenfunctions as the basis functions and $G(t)$, as defined in (21), is evaluated and optimized. In the present study, we use a spectral collocation method based on Chebyshev polynomials \citep{Schm01}. Figure \ref{fig:eig} shows the eigenvalue spectra for the Blasius boundary layer ($N = 0$) and for a boundary layer flow with uniform freestream shear ($N = 0.01$). The eigenvalue spectra are qualitatively similar for both the flows with a small shift between them. The straight line part in the respective spectra correspond to the discretised modes of the continuous spectrum \citep{Schm01}. The phase velocity of the disturbance is given by $c = \beta/\alpha = c_r + ic_i$. It can be noted that the phase velocity is greater than one for the uniform shear flow. This is due to the normalization by $U_{0}$ which is less than the total freestream velocity with uniform shear.

    \begin{figure}
 \unitlength=45.0mm
\centerline{
\includegraphics[width=1.9\unitlength,height=1.125\unitlength]{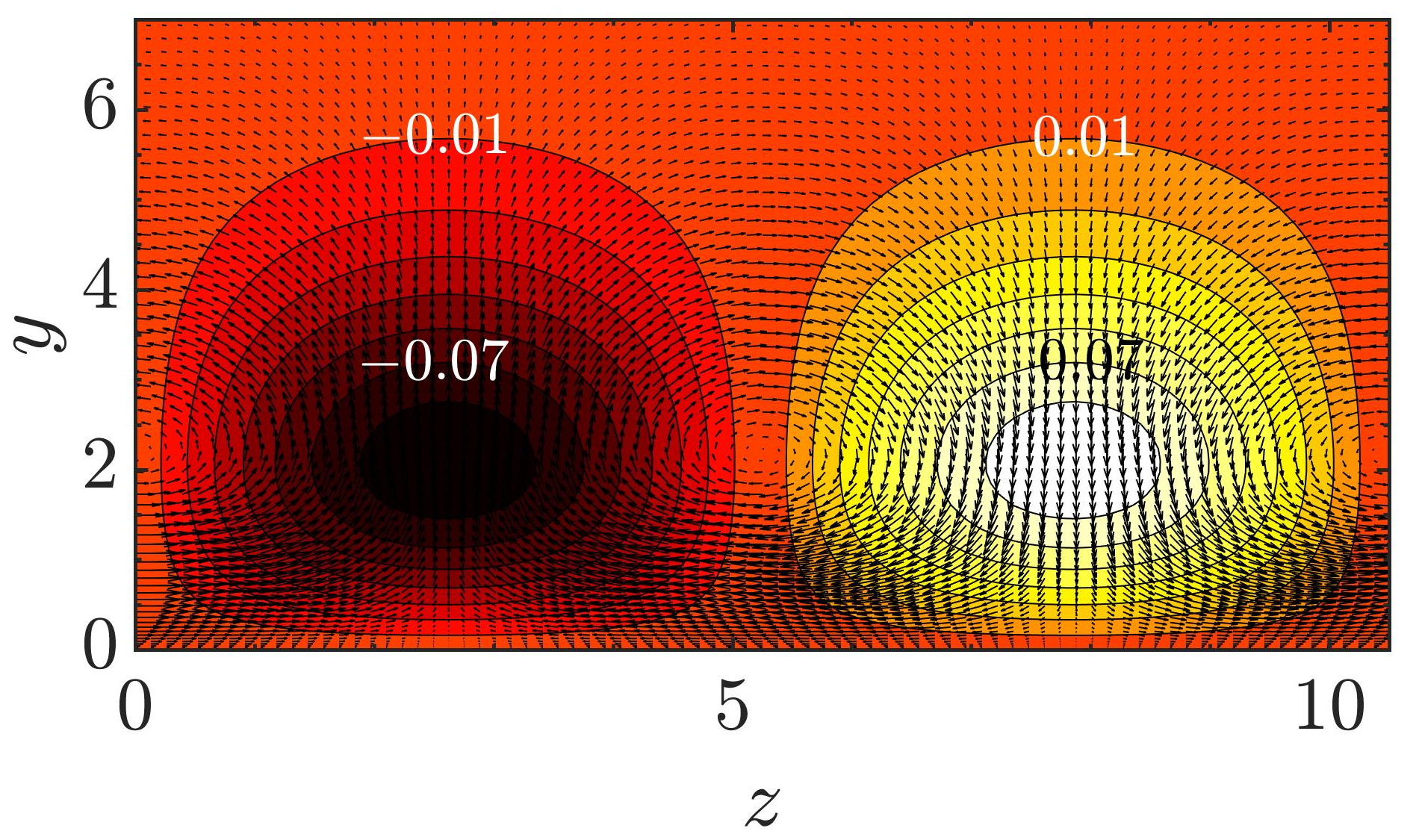}
}
\centerline{
\includegraphics[width=1.9\unitlength,height=1.125\unitlength]{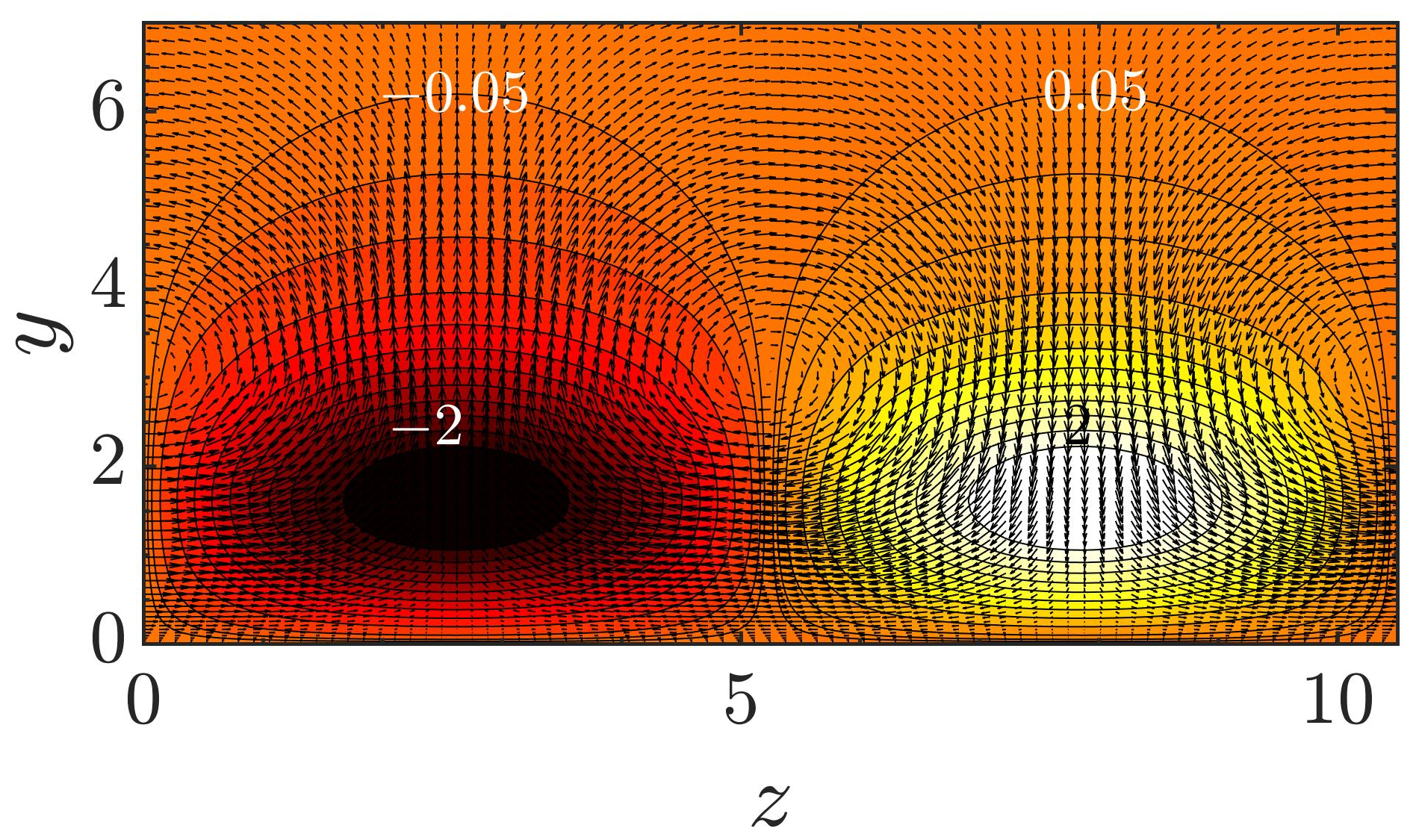}
}
\begin{picture}(0,0)
\put(-1.0,2.28){{(a)}}
\put(-1.0,1.14){{(b)}}
\end{picture}
 \caption{
Initial disturbance and its response in uniform freestream shear flows at $t$ = $t_{max}$ for $Re = 300$; $\alpha = 0$, $\beta$ $\approx$ 0.60. Vectors are of the spanwise and the wall normal disturbance velocities; line contours denote the streamwsie fluctuating velocity. (a) Optimal initial disturbance. (b) Optimal response.
}
 \label{fig:ini}
 \end{figure}
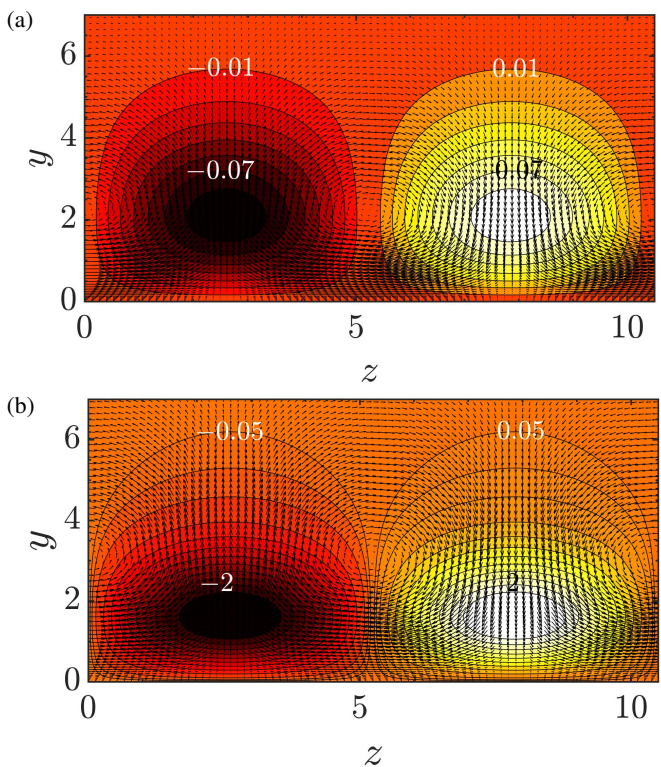

    \begin{figure}
 \unitlength=40.0mm
\centerline{
\includegraphics[width=2.05\unitlength,height=1.15\unitlength]{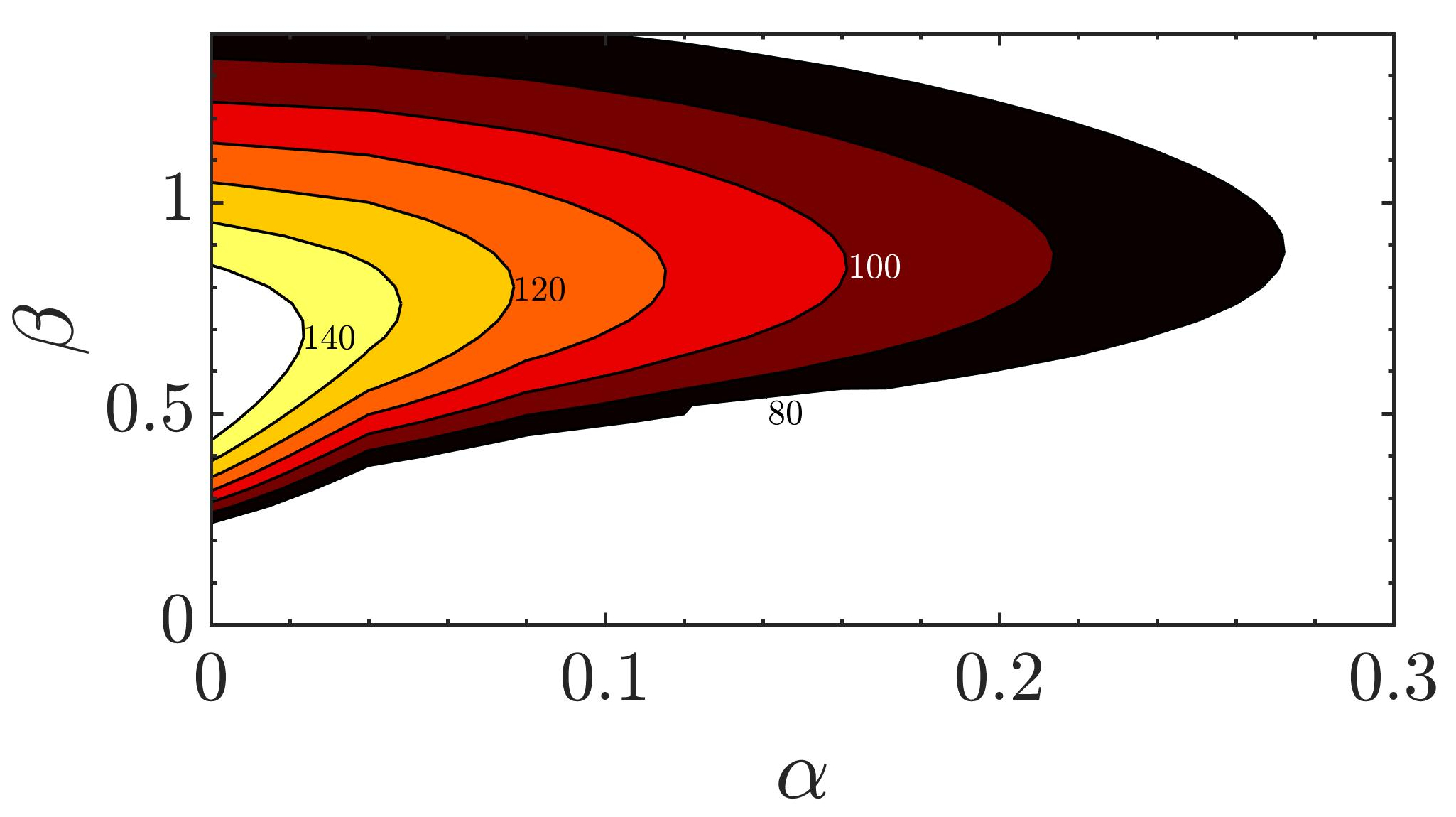}
}
\centerline{
\includegraphics[width=2.05\unitlength,height=1.15\unitlength]{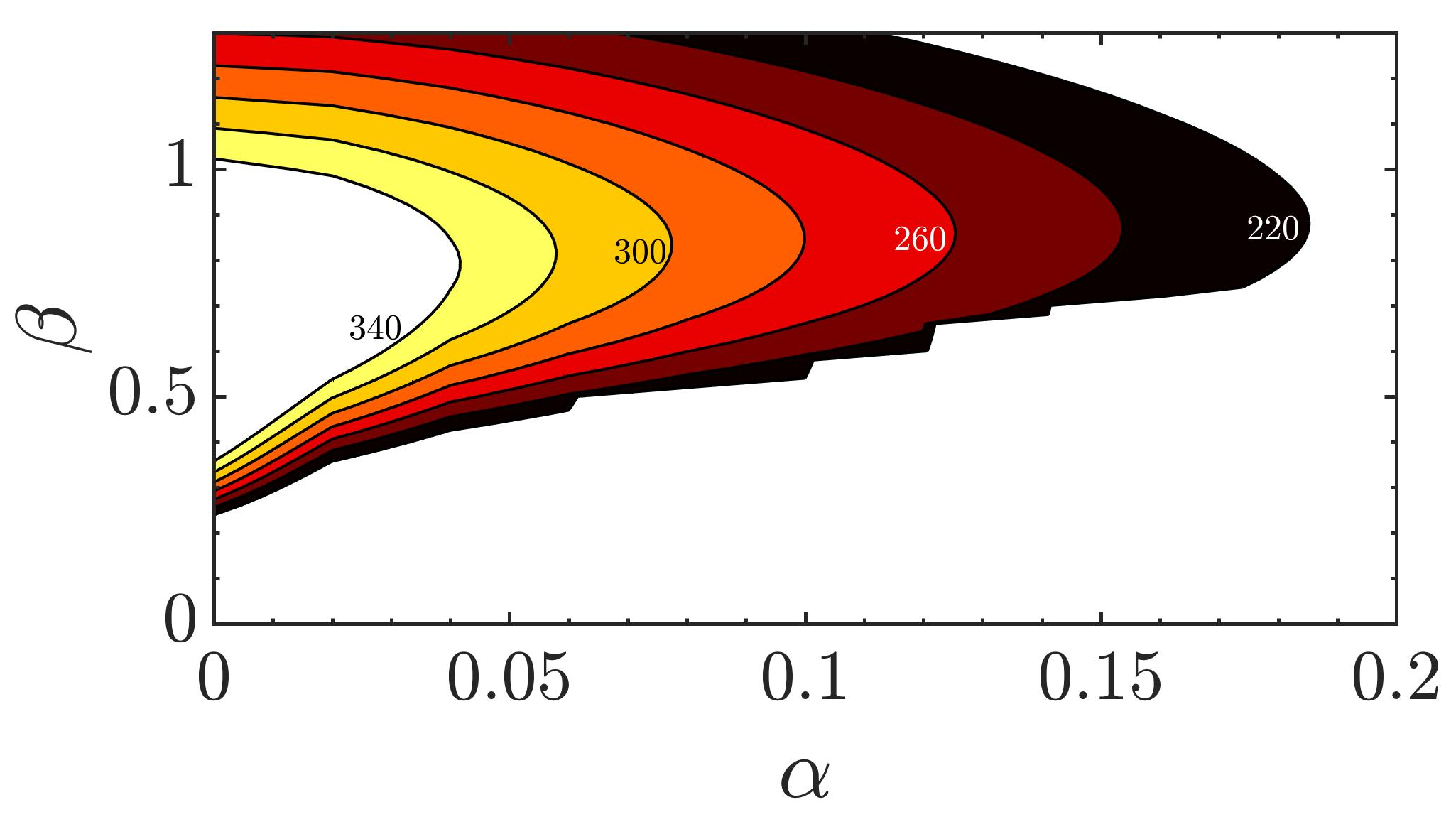}
}
\centerline{
\includegraphics[width=2.05\unitlength,height=1.15\unitlength]{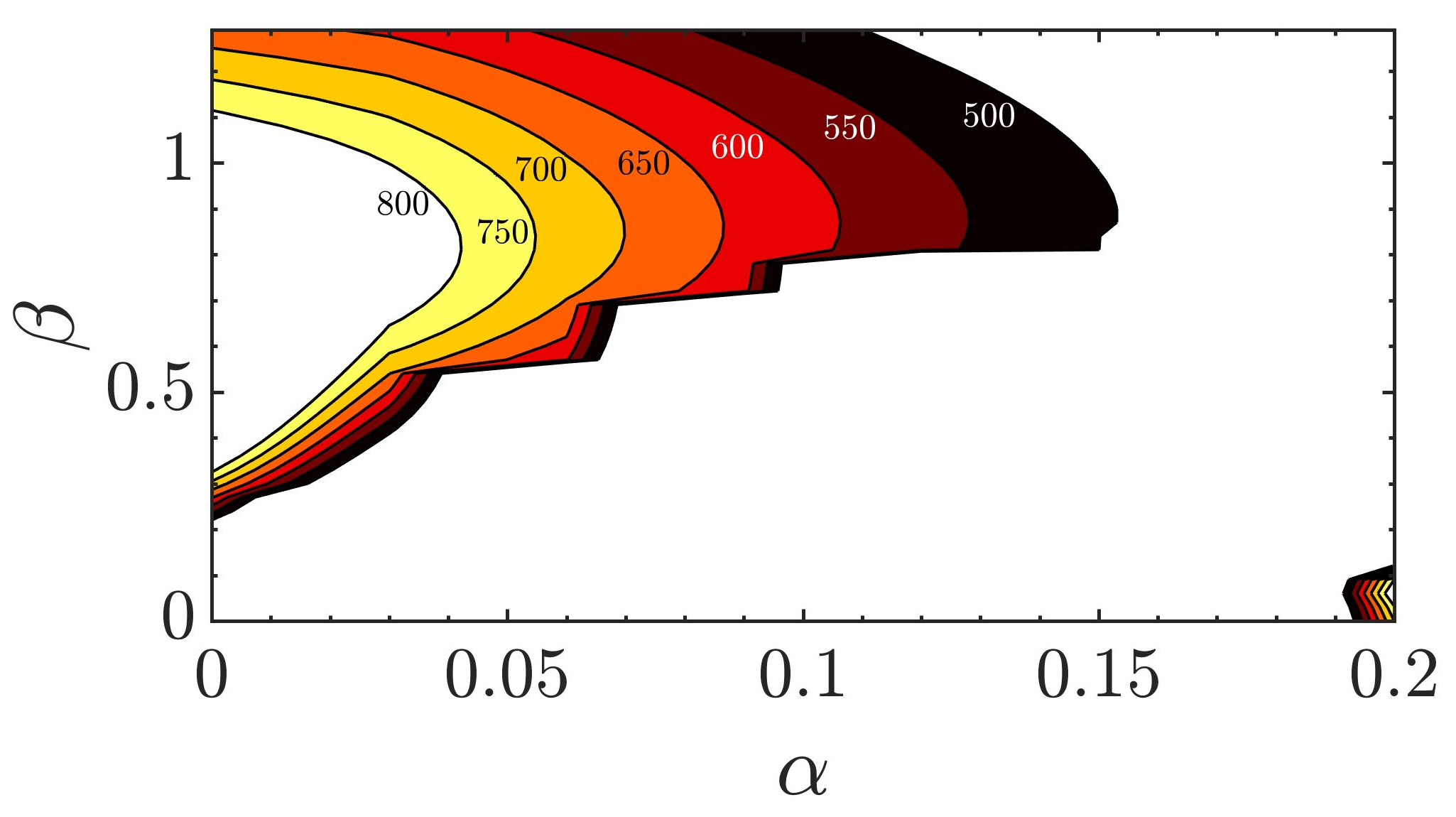}
}
\begin{picture}(0,0)
\put(-1.0,3.48){{(a)}}
\put(-1.0,2.32){{(b)}}
\put(-1.0,1.16){{(c)}}
\end{picture}
 \caption{
Contours of $G_{max}$($\alpha$, $\beta$, $Re$) in $\alpha$-$\beta$ plane for three different Reynolds numbers and $N = 0.01$. Parameter settings: (a) $Re = 300$; (b) $Re = 500$; (c) $Re = 800$. 
}
 \label{fig:diffr}
 \end{figure}
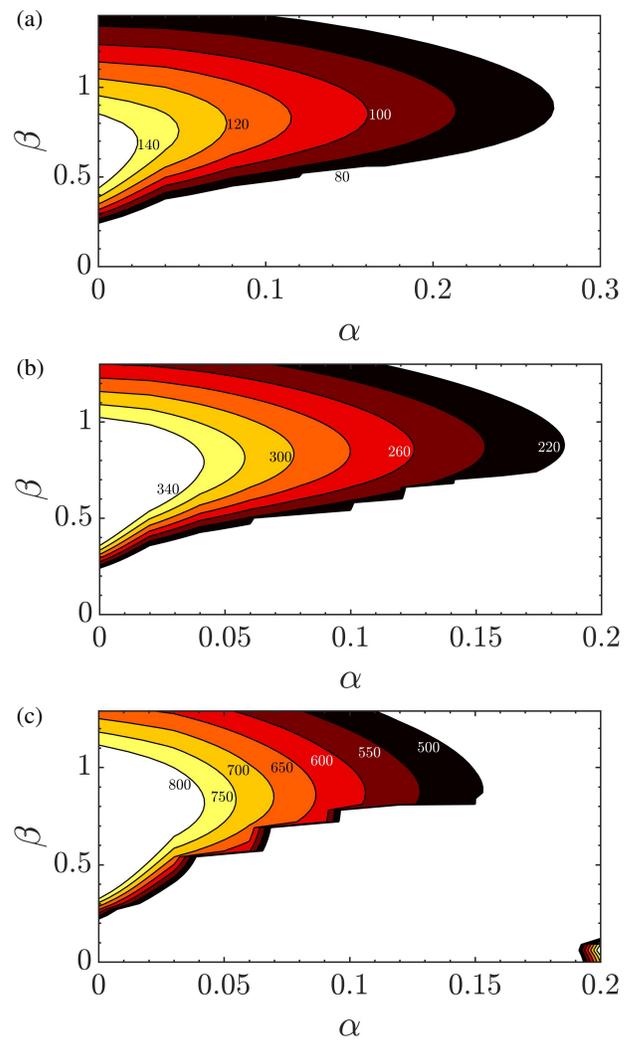 

Figure~\ref{fig:gt} shows the evolution of the energy amplification, $G(t)$, at a subcritical Reynolds number of $Re = 300$, for $N = 0$ and $N = 0.01$. For the Blasius boundary layer \citep{Butl92}, the optimum values of $\alpha$ and $\beta$ are $0$ and $0.65$, respectively, whereas they are $0$ and $0.60$, respectively, for the present uniform shear case. These values of $\alpha$ and $\beta$ are found to be optimum in the sense that an optimum energy amplification occurs at these values, as discussed below. A lower $\beta$ implies a higher spanwise wavelength than that for the Blasius flow. We may note that this is higher than the plane Couette flow \citep{Schm01} case $(\alpha = 35/Re, \beta = 1.6$; $Re$ appropriate to this flow). We see that there is significant transient growth in a boundary layer flow with uniform freestream shear as well. It can be seen that in the initial phase, the growth of $G(t)$ is the same for both the uniform shear and Blasius flow, and is unaffected by the freestream shear. However, a higher peak transient amplification is observed for the freestream shear flow.

For $N = 0.01$, the optimal initial disturbances which induce the maximum growth in the boundary layer, and the optimal response are shown in Fig.~\ref{fig:ini}. We see that the optimal initial disturbances which trigger the maximum growth are the streamwise vortices. These are similar to those in other shear flows \citep[][]{Ande99,Butl92,Corb00}. We can also note from the contour levels that the maxima of the optimal response is orders of magnitude larger than that of the optimal initial disturbance representative of the algebraic growth of energy. Thus the emergence of strong streamwise velocity streaks in Fig.~\ref{fig:ini}(b) may be explained as the result of the initial streamwise vortex \cite{Henningson2006}.

   \begin{figure}
 \unitlength=40.0mm
\centerline{
\includegraphics[width=2.07\unitlength,height=1.2\unitlength]{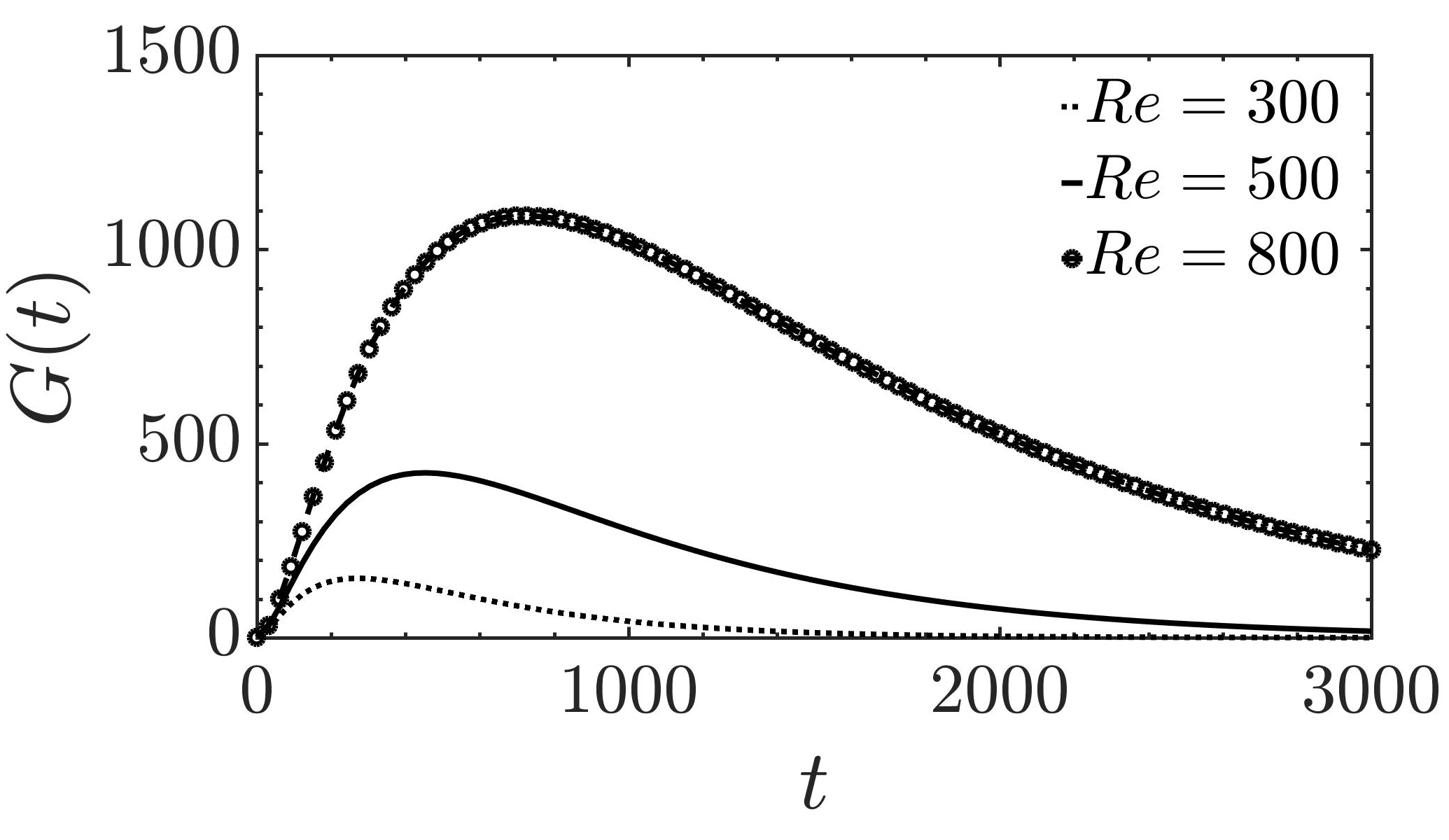}
}
 \caption{
Temporal variation of $G(t)$ at three different Reynolds numbers. Parameter settings: $N = 0.01$, $\alpha = 0$ and $\beta = 0.60$.
}
 \label{fig:gtus}
 \end{figure}

For the boundary layer flow corresponding to $N = 0.01$, $G_{max}$ is found to be optimum at $\alpha = 0$ and $\beta \approx 0.60$. This may be clearly seen in Fig.~\ref{fig:diffr}, where the level curves of $G_{max}$ are shown in the $\alpha$-$\beta$ plane for three different Reynolds numbers. The peak value of $G_{max}$ in the $\alpha$-$\beta$ plane corresponds to the optimal energy growth, $G_{opt}$. One may also notice that the value of $G_{opt}$ increases with increasing Reynolds number. The effect of increasing Reynolds number in the transient growth amplification is shown in Fig.~\ref{fig:gtus}. For a particular value of $N$, indeed $G_{opt}$ depends on the Reynolds number. We may also note that the contour lines in the lower branch of each frame in Fig.~\ref{fig:diffr} are not smooth. The kink on the constant growth rate curve in the Orr--Sommerfeld solutions \cite{Bera05} may be a reason for this effect.

    \begin{figure}
 \unitlength=40.0mm
\centerline{
\includegraphics[width=2.08\unitlength,height=1.16\unitlength]{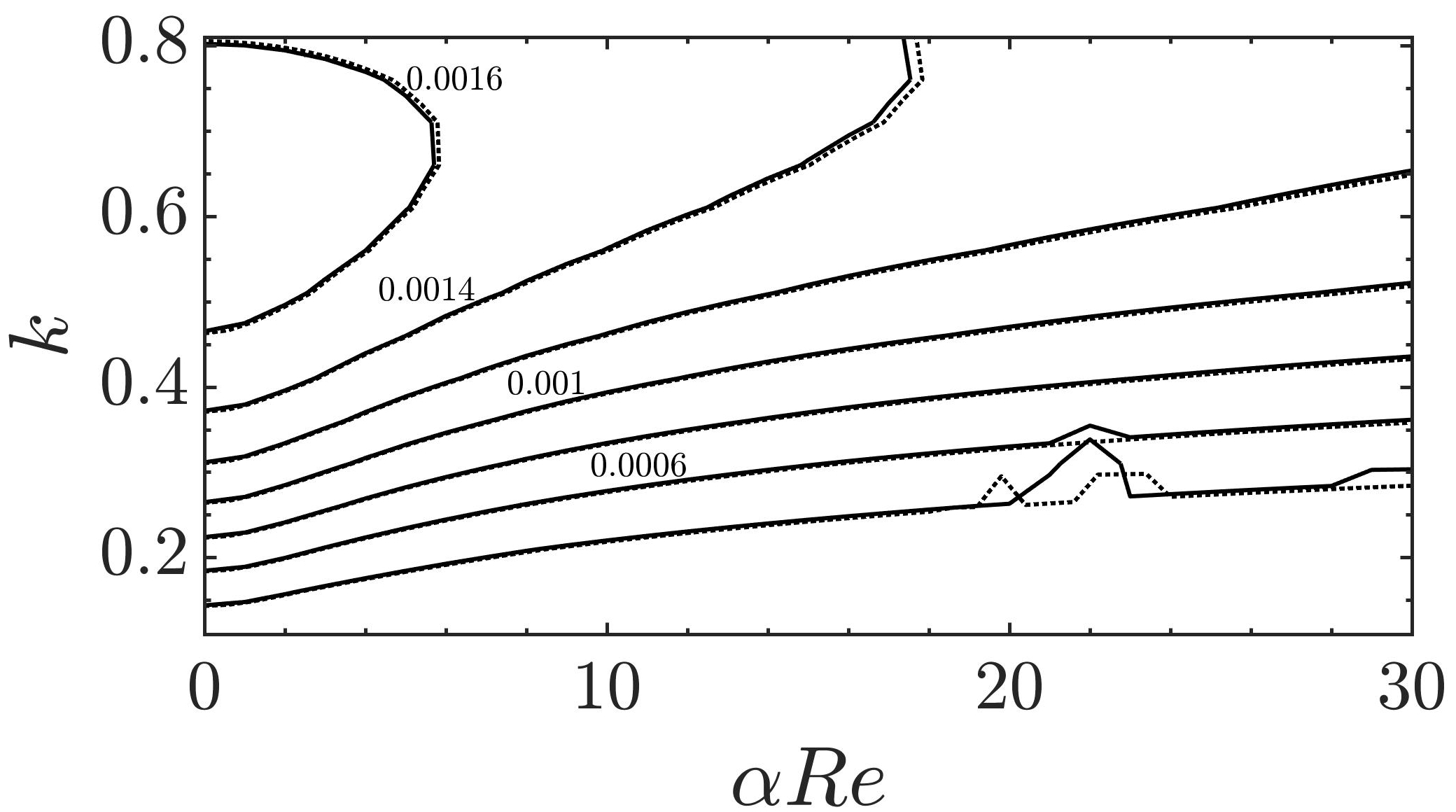}
}
 \caption{
Contours of $k^{2}G_{max}(\alpha, \beta, Re)/\beta^{2}Re^{2}$ in $k$ and $\alpha$-$Re$ plane. Solid line:  $Re = 500$; dashed line: $Re = 300$. Parameter settings: $N = 0.01$.
}
 \label{fig:kar}
 \end{figure}

   \begin{figure}
 \unitlength=40.0mm
\centerline{
\includegraphics[width=2.0\unitlength,height=1.2\unitlength]{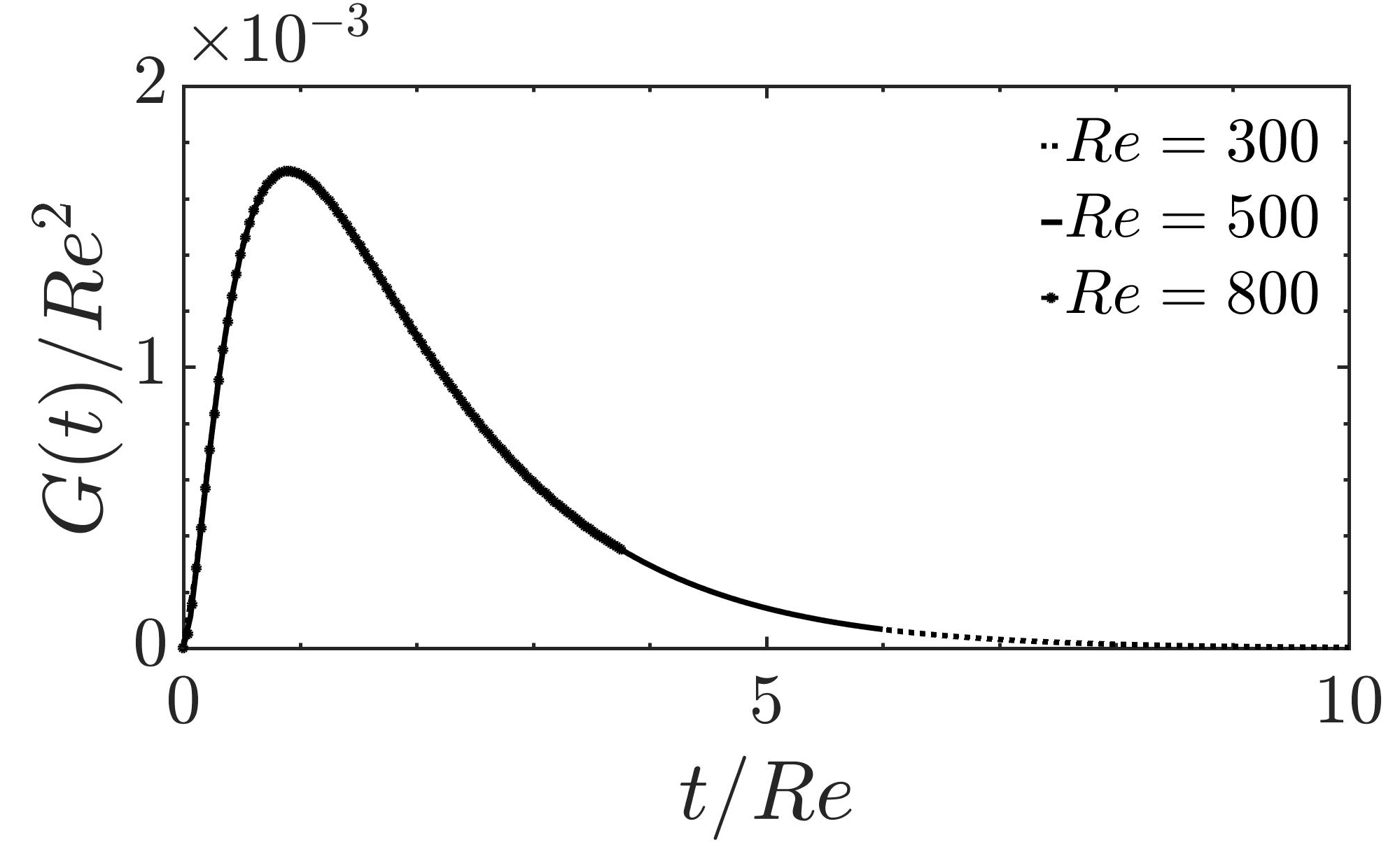}
}
 \caption{Curves of maximum optimal amplification for different $Re$ with $G(t)$ rescaled by $Re^{2}$ and $t$ by $Re$. Parameter settings: $N=0.01$, $\alpha = 0$, $\beta = 0.60$.
}
 \label{fig:sca}
 \end{figure}

      \begin{figure}
 \unitlength=40.0mm
\centerline{
\includegraphics[width=2.05\unitlength,height=1.15\unitlength]{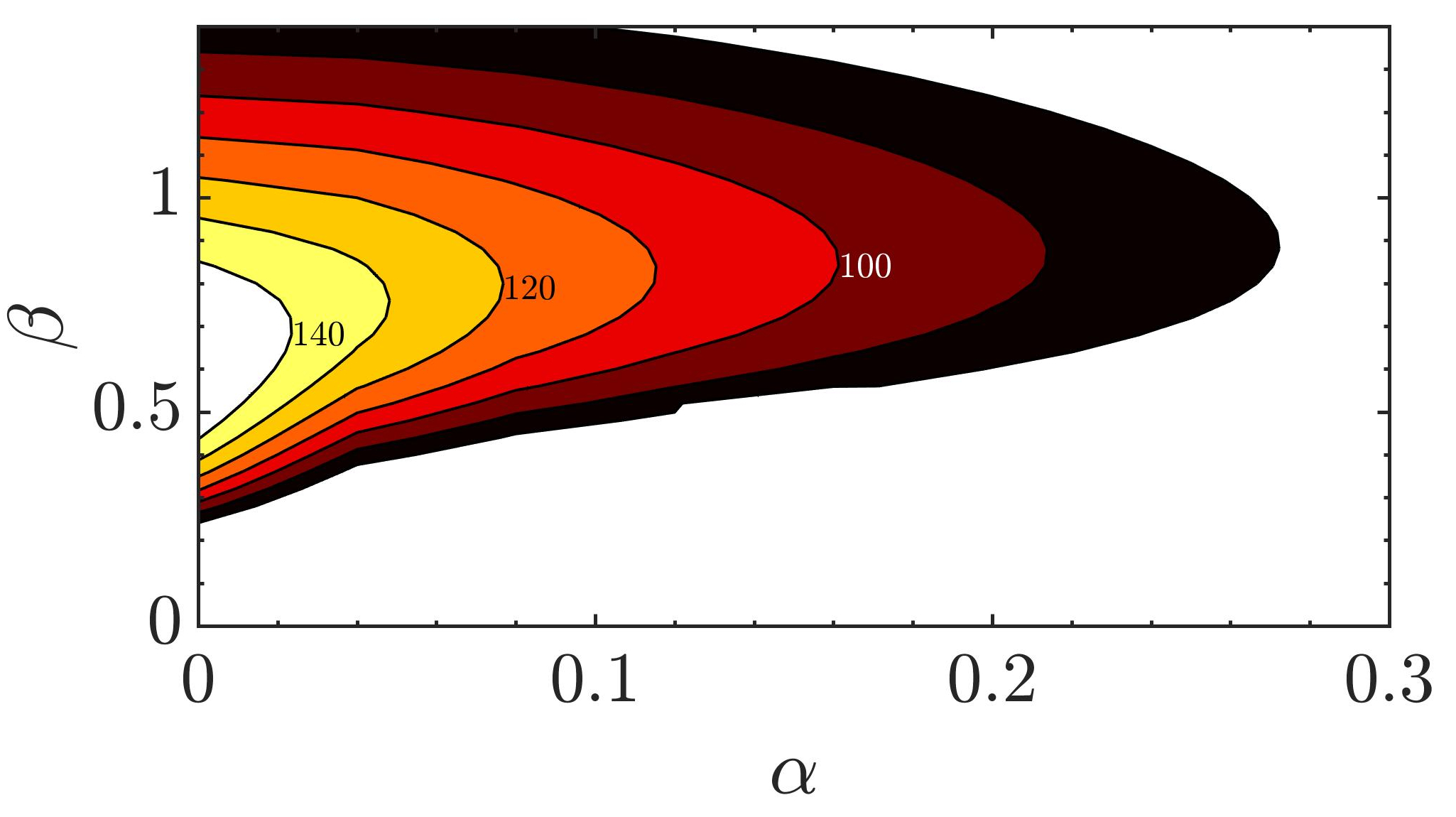}
}
\centerline{
\includegraphics[width=2.05\unitlength,height=1.15\unitlength]{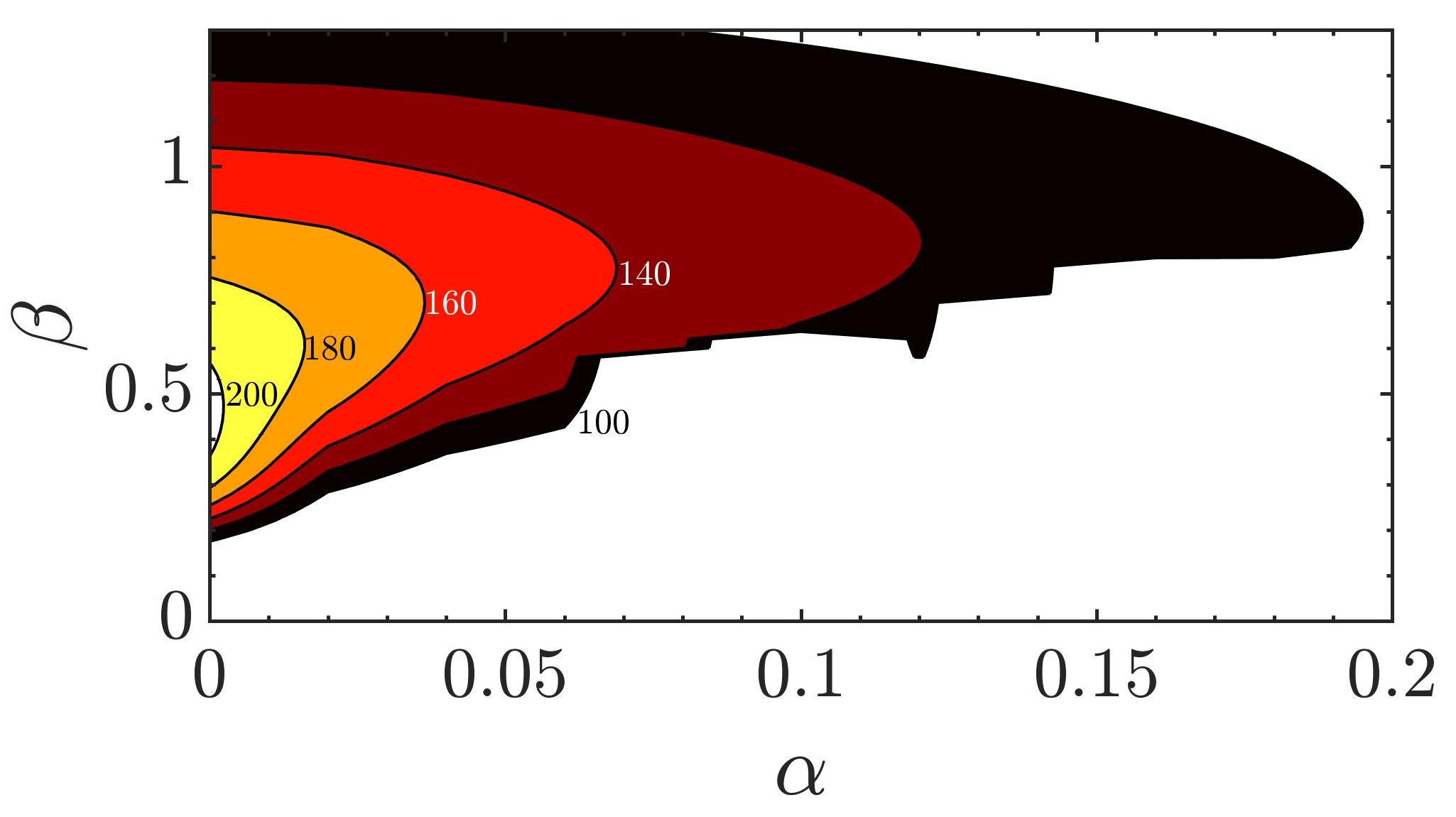}
}
\centerline{
\includegraphics[width=2.05\unitlength,height=1.15\unitlength]{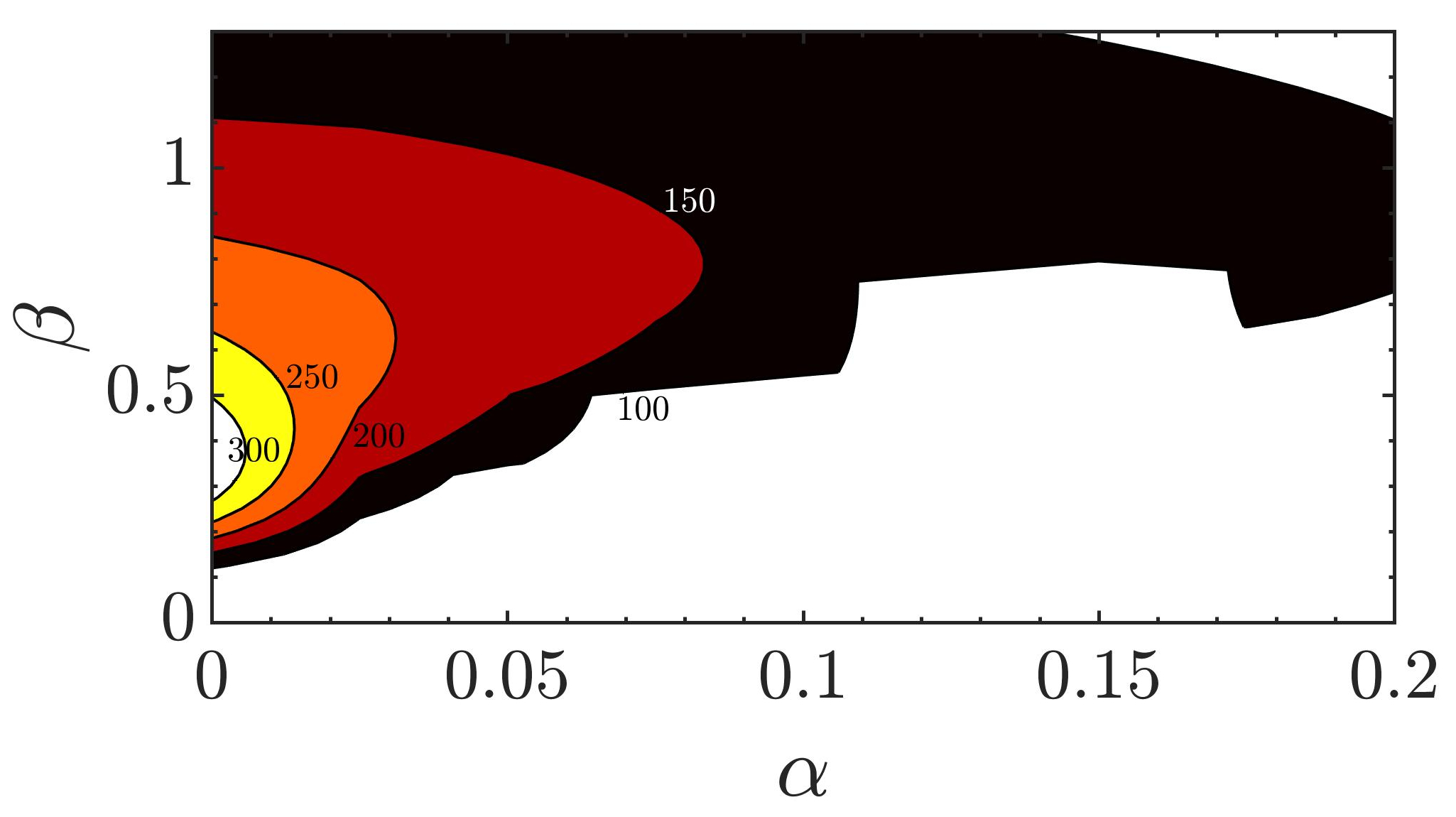}
}
\begin{picture}(0,0)
\put(-1.0,3.48){{(a)}}
\put(-1.0,2.32){{(b)}}
\put(-1.0,1.16){{(c)}}
\end{picture}
 \caption{
Contours of $G_{max}(\alpha,\,\beta,\,Re$) in the $\alpha$-$\beta$ plane at $Re = 300$ for different values of $N$. Parameter settings: (a) $N = 0.01$, (b) $N = 0.03$, (c) $N = 0.05$.
}
 \label{fig:diffn}
 \end{figure}
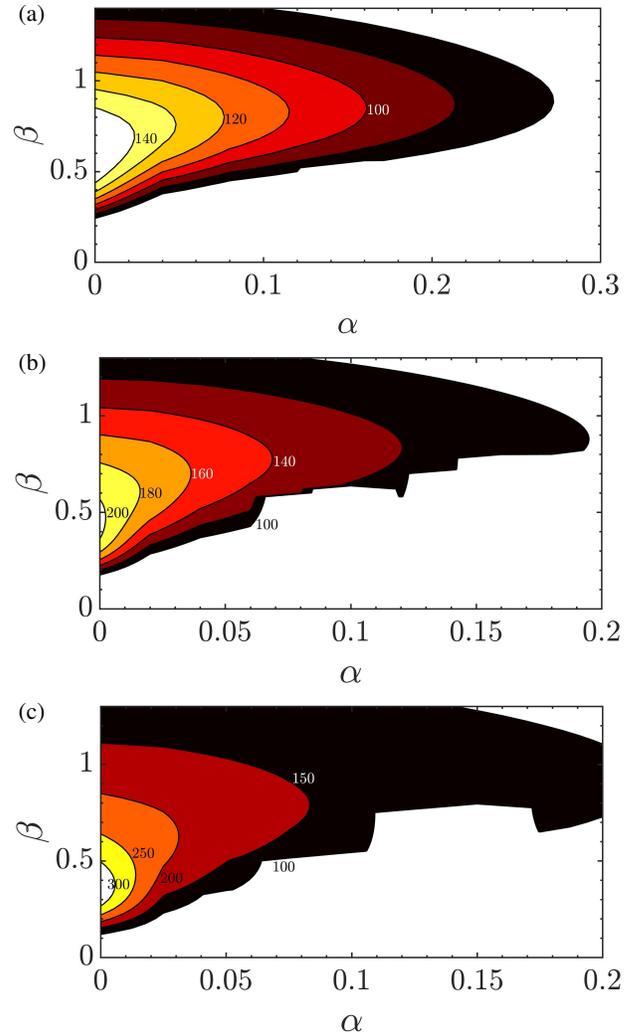

   Since an estimation of $G_{max}$($\alpha$,\,$\beta$,\,$Re$) in the three-dimensional parameter space ($\alpha$,\,$\beta$,\,$Re$) is computationally expensive, Reddy \& Henningson \cite{Redd93} derived a scaling relation in the
two parameter $(k,\alpha Re)$ space for $G_{max}$($\alpha$,\,$\beta$,\,$Re$) given by
\begin{equation}
G_{max}(\alpha,\beta,Re) \approx
\frac{\beta^{2}Re^{2}}{k^{2}}H_{2}(k,\alpha Re),
\end{equation}
for some function $H_{2}$. This scaling relation holds good for Couette and Poiseuille flows more accurately, for $\alpha Re \rightarrow 0$ \cite{Redd93}. As shown in Fig.~\ref{fig:kar}, where the level curves of $k^{2}G_{max}(\alpha,\,\beta,\,Re)/\beta^{2}Re^{2}$ are shown for two Reynolds numbers, this scaling relation holds good even for a boundary layer developing under a freestream with uniform shear. This quadratic scaling of energy is clearly seen in Fig.~\ref{fig:sca}, where $G(t)$ and $t$ are scaled with $Re^{2}$ and $Re$, respectively. For $N = 0.01$, the optimum value of $G_{max}$ in the $\alpha$-$\beta$ plane scales as $1.69\times10^{-3}Re^{2}$, whereas for the Blasius boundary layer \cite{Butl92,Schm01}, the same scales as $1.50\times10^{-3}Re^{2}$.

\onecolumngrid

   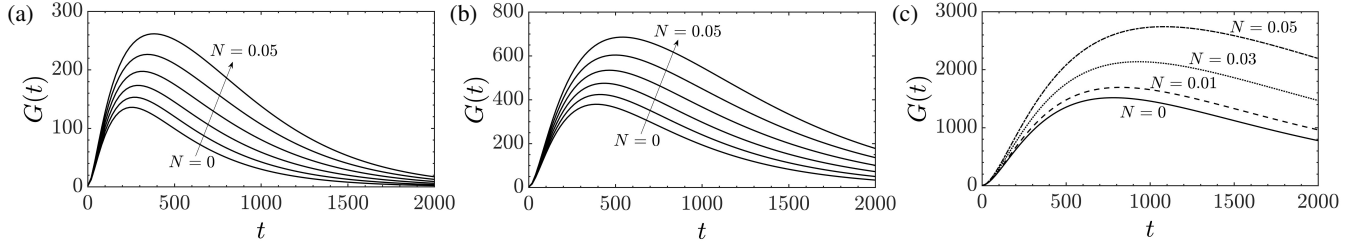
\begin{figure}[h]
 \unitlength=28.0mm
\centerline{
\includegraphics[width=2.07\unitlength,height=1.2\unitlength]{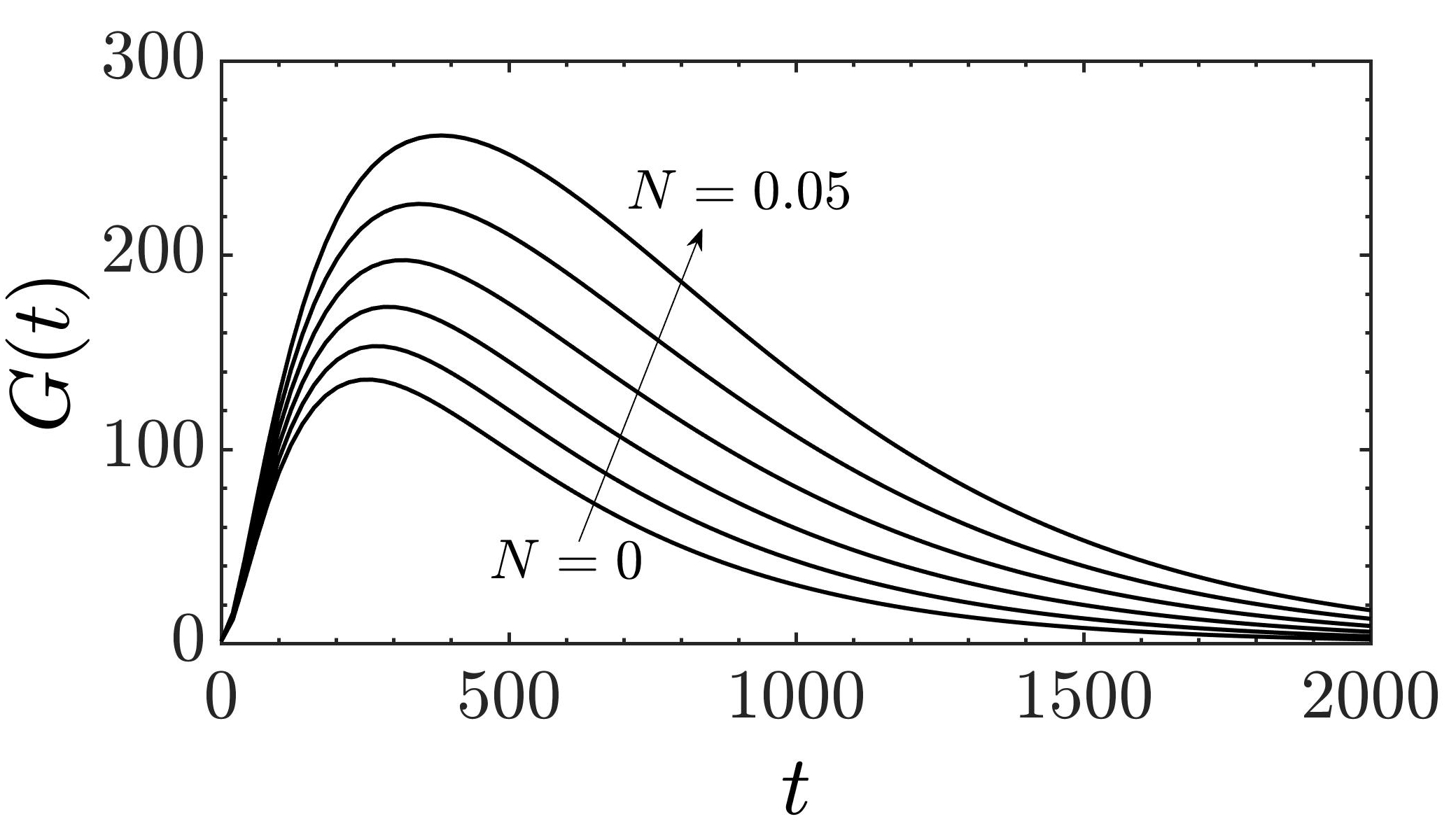}
\includegraphics[width=2.07\unitlength,height=1.2\unitlength]{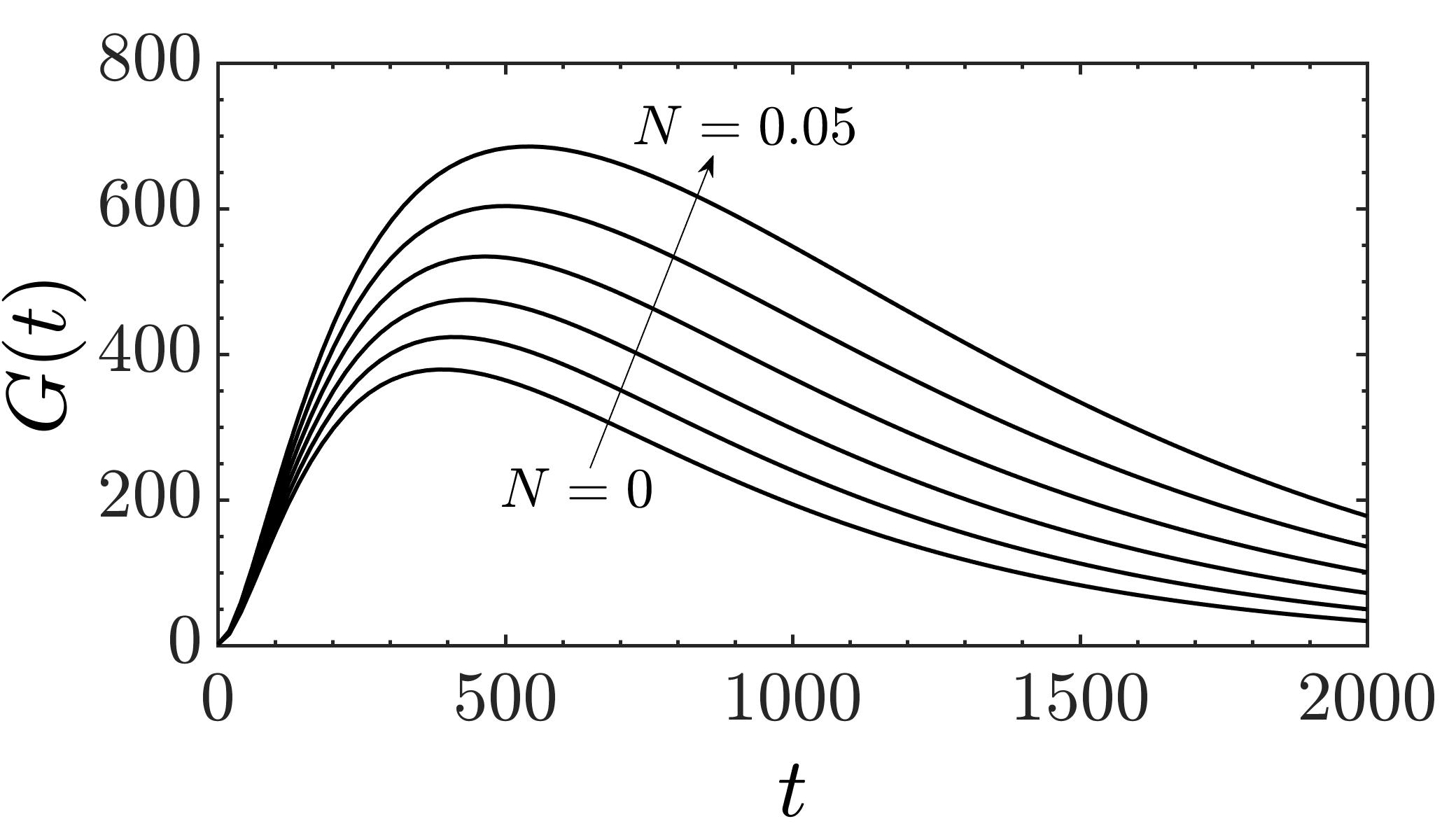}
\includegraphics[width=2.07\unitlength,height=1.2\unitlength]{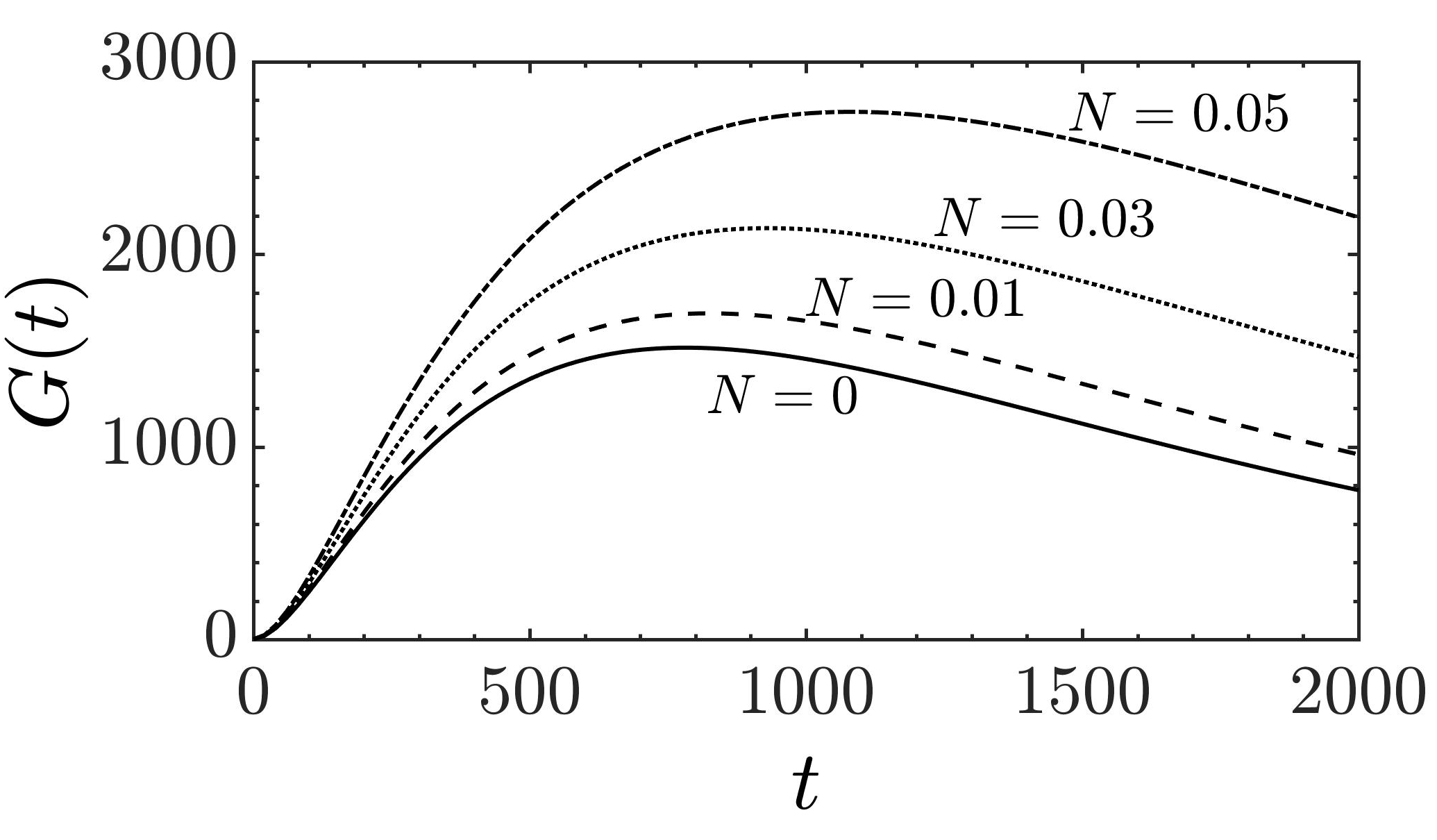}
}
\begin{picture}(0,0)
\put(-3.2,1.2){{(a)}}
\put(-1.1,1.2){{(b)}}
\put(1.0,1.2){{(c)}}
\end{picture}
 \caption{
Maximum amplification, $G(t)$, for different values of $N$. Parameter settings: (a) $Re = 300, \alpha = 0.0, \beta = 0.60$.
(b) $Re = 500, \alpha = 0.0, \beta = 0.65$, (c) $Re = 1000, \alpha = 0.0, \beta = 0.65$.
}
 \label{fig:gtN}
 \end{figure}

  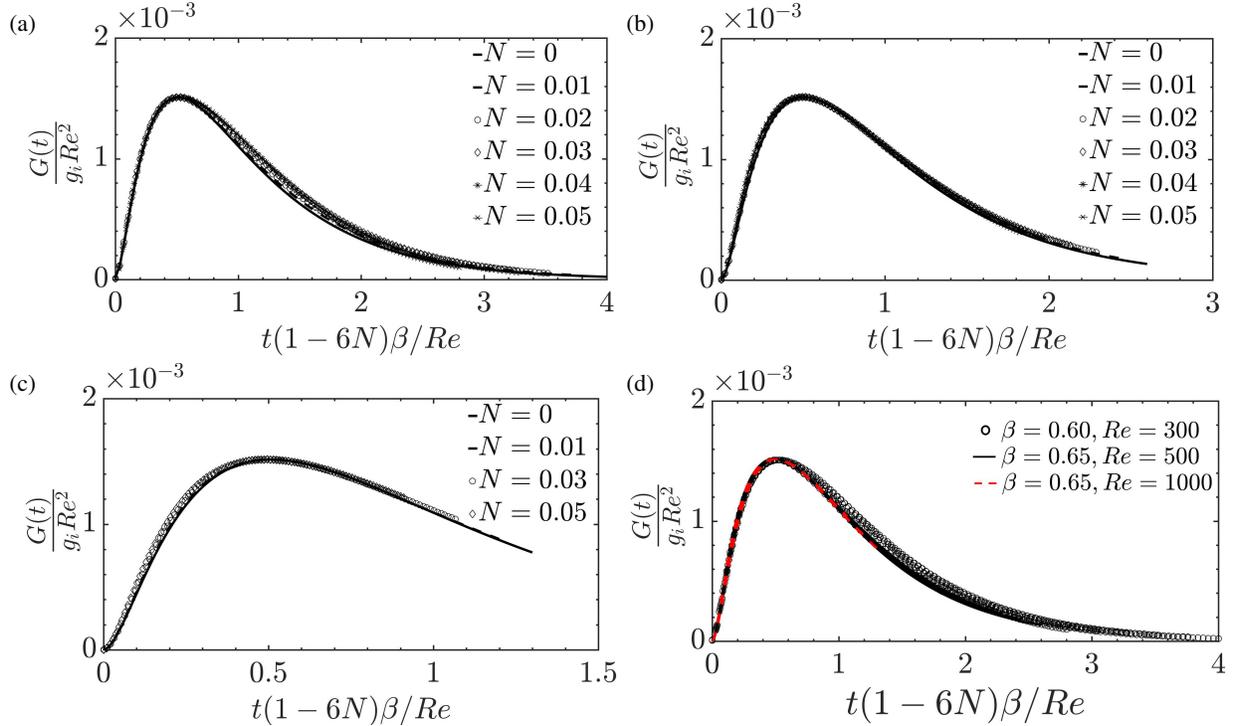
\begin{figure}[h]
 \unitlength=40.0mm
\centerline{
\includegraphics[width=2.0\unitlength,height=1.2\unitlength]{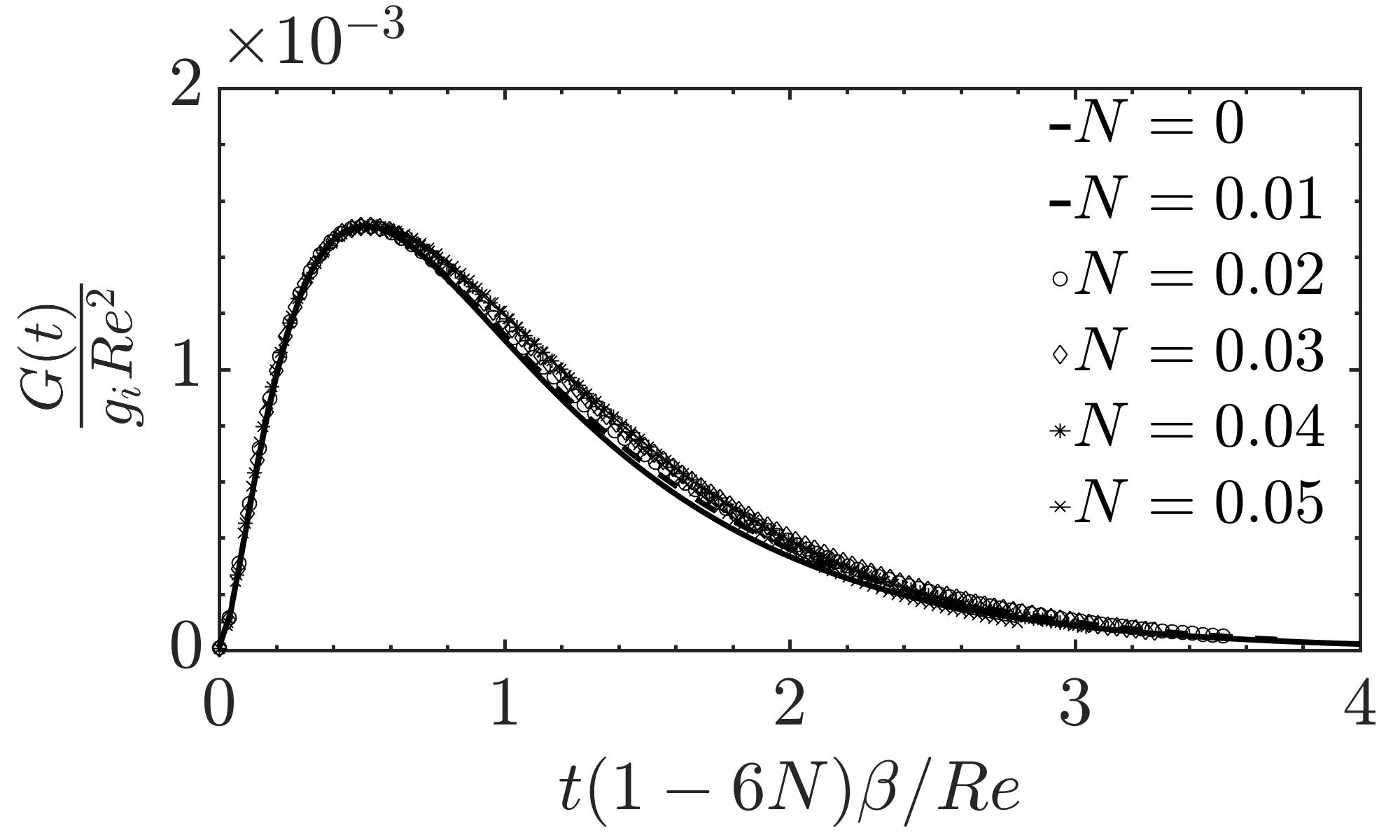}
\includegraphics[width=2.0\unitlength,height=1.2\unitlength]{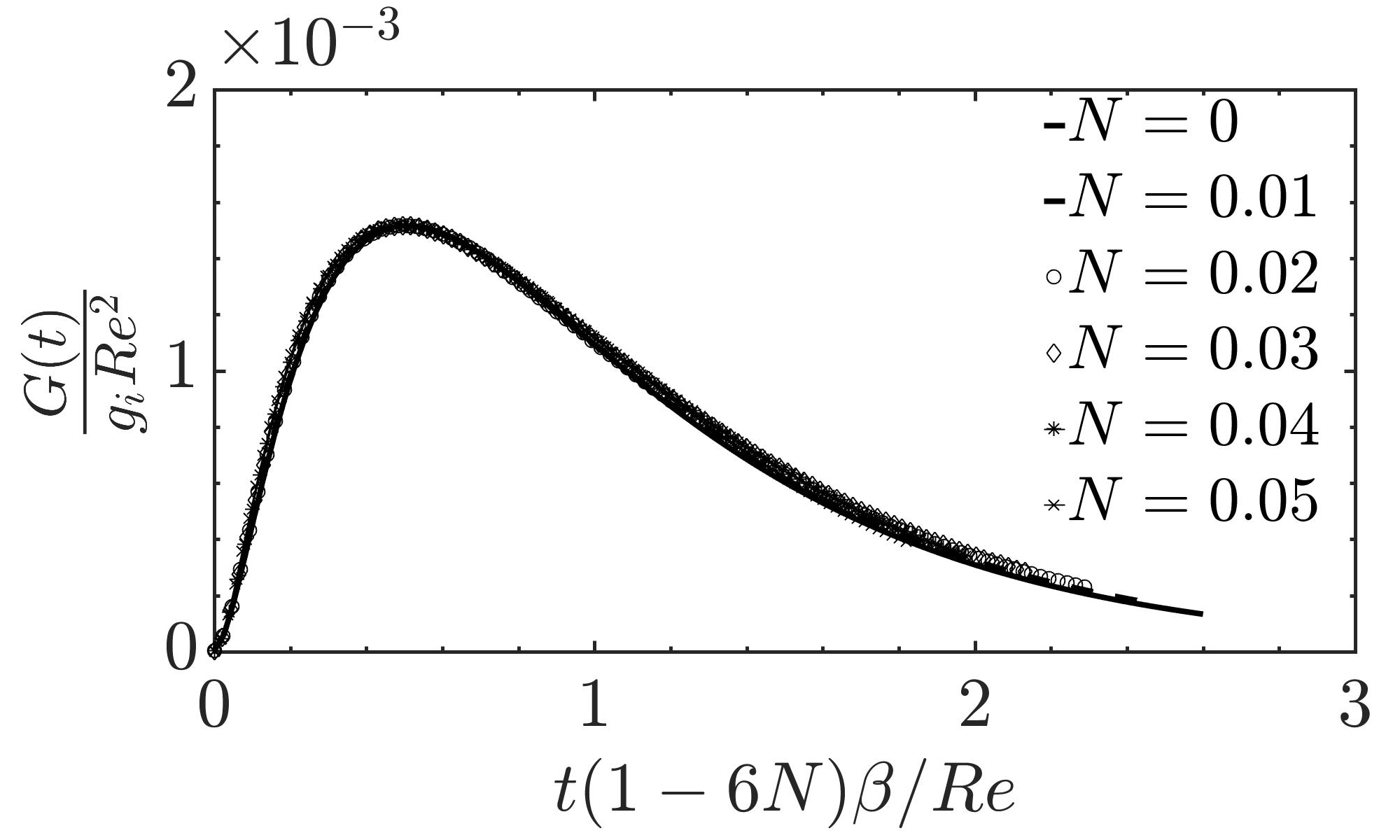}
}
\centerline{
\includegraphics[width=2.0\unitlength,height=1.2\unitlength]{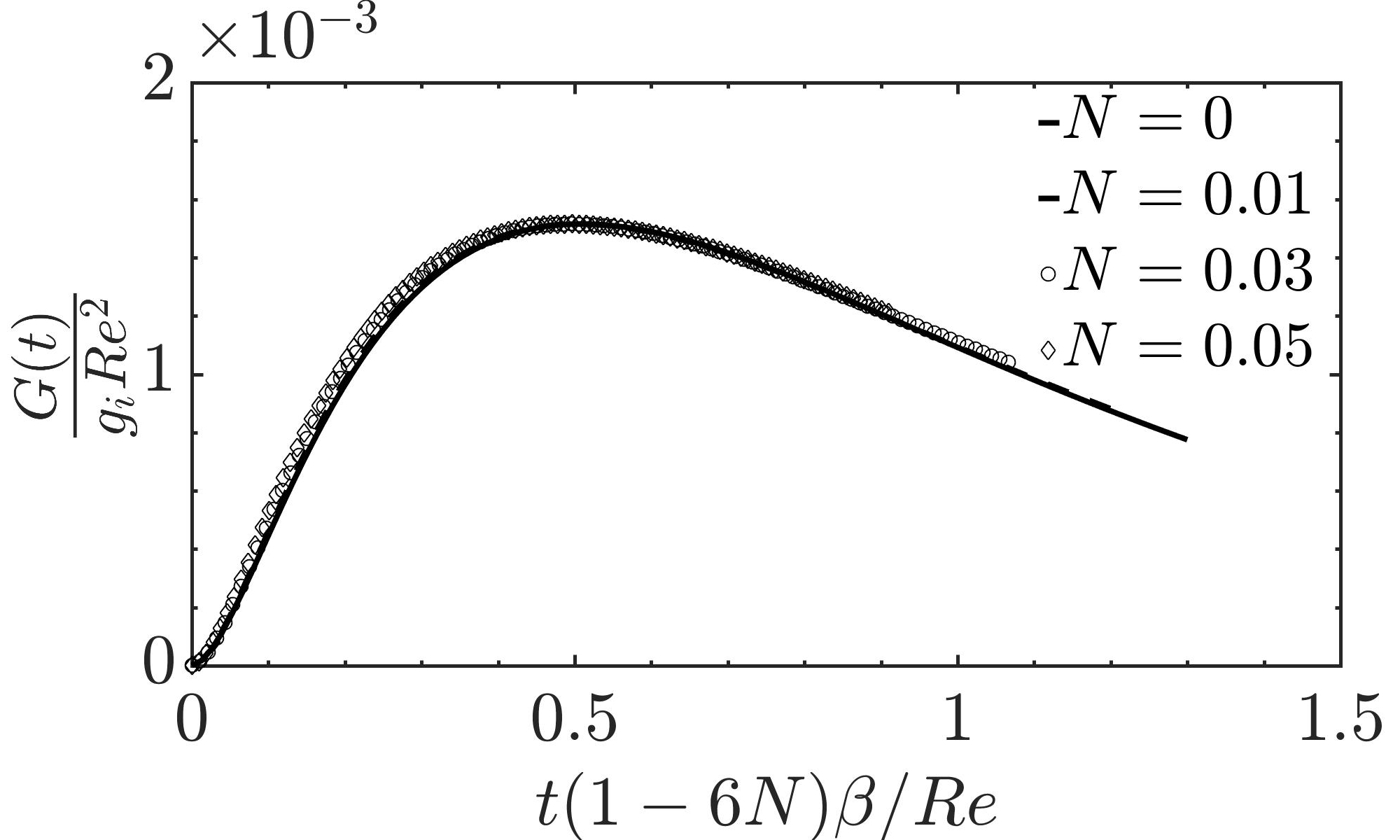}
\includegraphics[width=2.0\unitlength,height=1.2\unitlength]{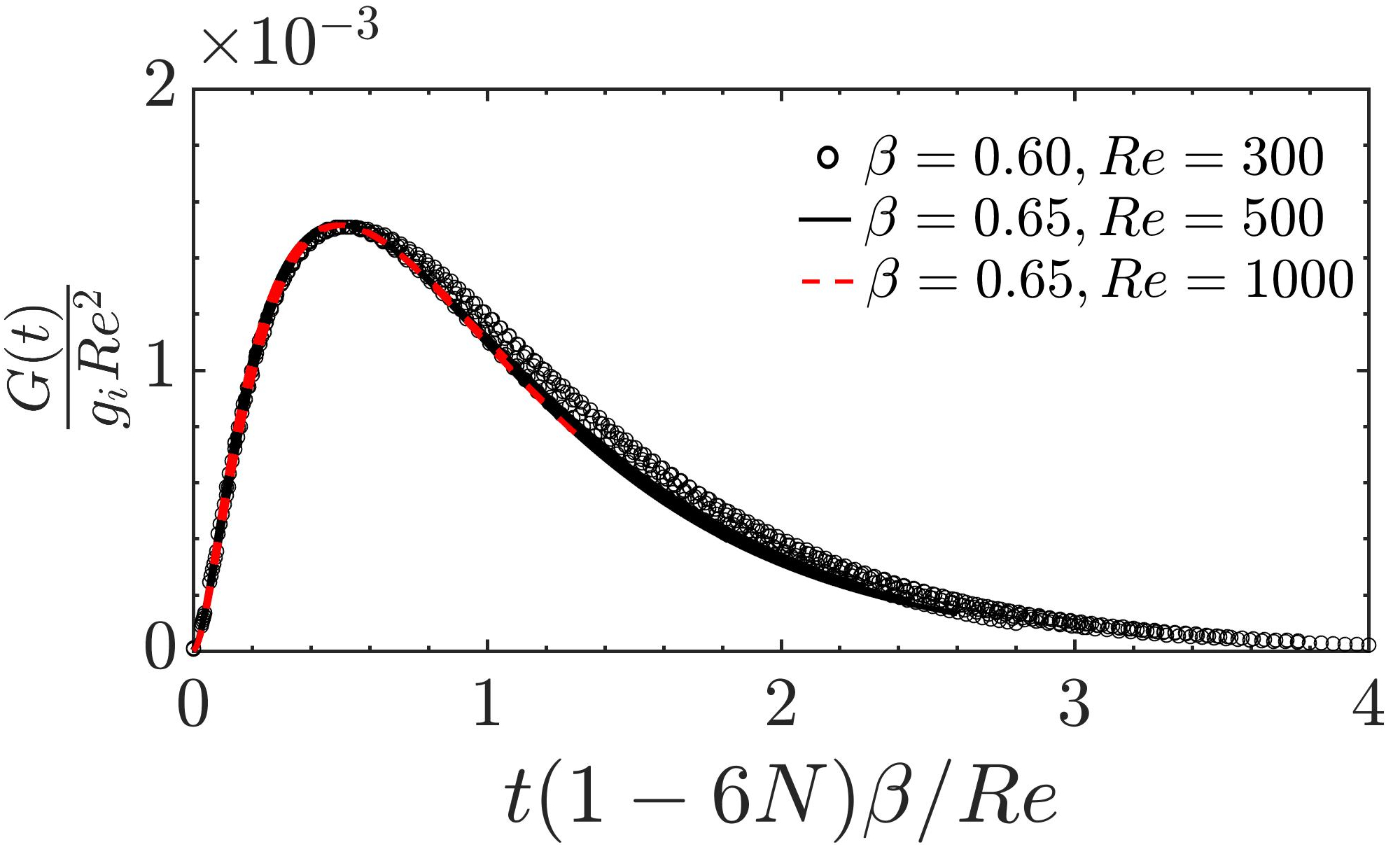}
}
\begin{picture}(0,0)
\put(-2.05,2.4){{(a)}}
\put(0.0,2.4){{(b)}}
\put(-2.05,1.2){{(c)}}
\put(0.0,1.2){{(d)}}
\end{picture}
 \caption{
Rescaled $G(t)$ curves are shown in panels (a,b,c) for the same parameter settings as in Fig.\ref{fig:gtN}. The scaling law holds well for all the different cases, which is summarized in panel (d).
}
 \label{fig:gtNscal}
 \end{figure}
 
\twocolumngrid

  \begin{figure}
 \unitlength=40.0mm
\centerline{
\includegraphics[width=2.0\unitlength,height=1.21\unitlength]{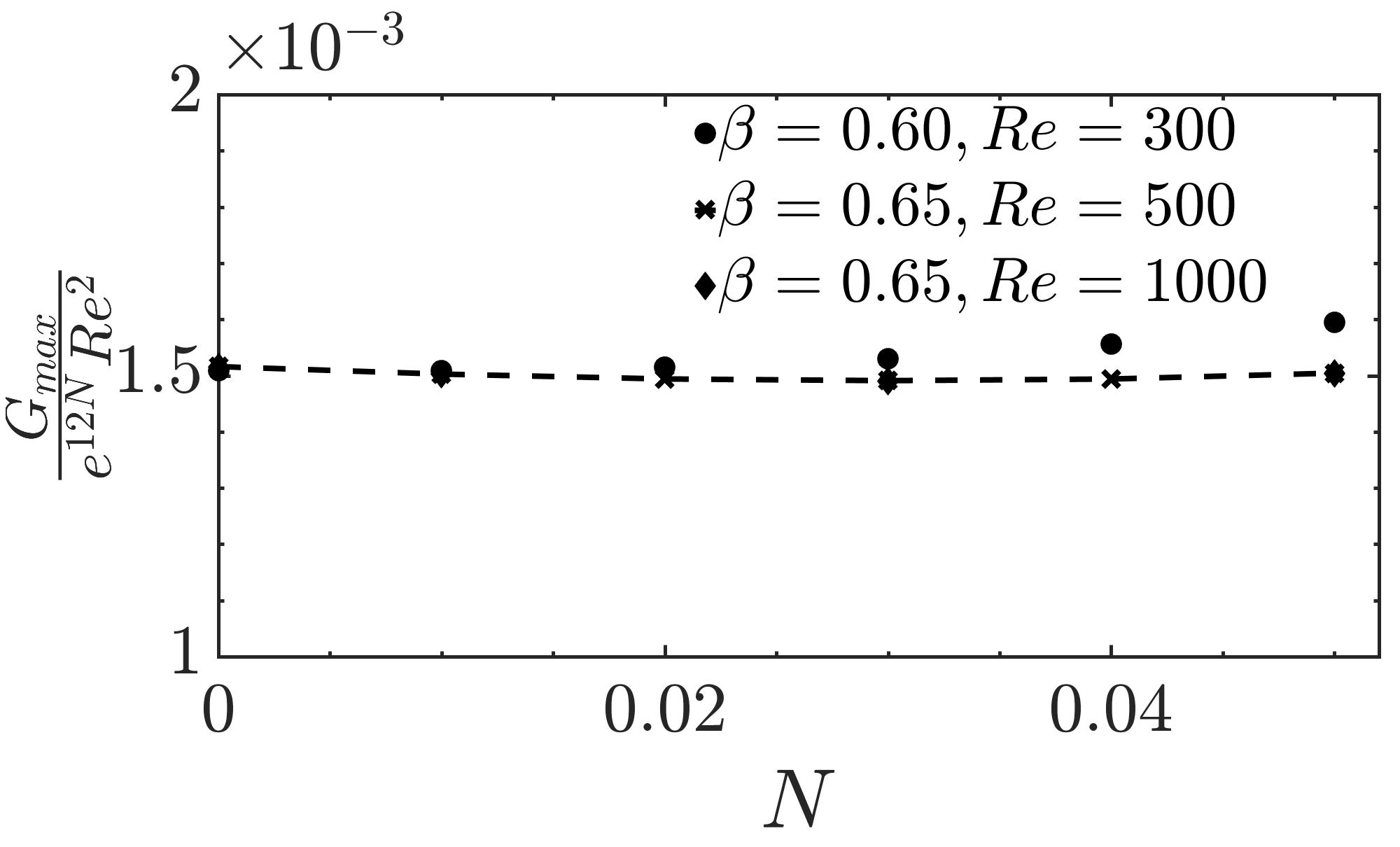}
}
 \caption{
Maximum optimal amplification, $G_{max}$, rescaled by $e^{12N} Re^{2}$, as a function of the freestream shear $N$ for the different cases considered in the present study.
}
 \label{fig:gtNmax}
 \end{figure}

 So far, we have discussed the results for $N = 0.01$. We now vary the freestream shear at a fixed value of the Reynolds number.  Figure \ref{fig:diffn} shows the contours of $G_{max}(\alpha,\,\beta,\,Re$) for $N = 0.01$, $0.03$ and $0.05$ at $Re = 300$. The optimum value of $G_{max}(\alpha,\,\beta,\,Re$) in the $\alpha$-$\beta$ plane is found to increase with increasing $N$, at a given Reynolds number. 
 With increasing $N$, the peak value of $G_{max}(\alpha,\,\beta,\,Re$) in the $\alpha$-$\beta$ plane occurs at a lower value of $\beta$, but at $\alpha = 0$. 
 Specifically, the optimum growth for $N = 0.03$ occurs at $\alpha = 0$ and $\beta$ $\approx$ $0.44$, and at $\alpha = 0$ and $\beta = 0.36$ for $N = 0.05$.

Figure \ref{fig:gtN} summarizes the effect of increasing freestream shear on the transient energy growth at three different values of the Reynolds number. It can be seen that the maximum amplification increases with increasing $N$. Though the linear stability analysis has shown that the freestream shear stabilises the flow \cite{Bera05}, such an increase in amplfication can lead to a bypass transition scenario \cite{balamurugan2017experiments1}. This is in congruence with the experimental observations reported in \cite{balamurugan2017experiments1} wherein they found the wall normal distribution of the normalized $u_{\textrm{rms}}$ profile to follow the non-modal theory of Luchini \cite{luchini2000reynolds}. In addition, it can be seen that the peak transient amplification is delayed with increasing $N$ and $Re$.

The rescaled $G(t)$ curves are shown in Fig. \ref{fig:gtNscal}, where $G(t)$ is scaled as $g_{i}Re^{2}$, and $t$ by $(1-6N)\beta/Re$. Here $g_i$ denotes the ratio of $G_{max}$ for the corresponding value of $N$ normalized by $G_{max}$ at $N = 0$, which is the Blasius flow. 
The scaling law holds well for all the different cases considered in the present study, which is evident from Fig. \ref{fig:gtNscal}(d). As remarked earlier, the time instant at which the maximum optimal growth is attained increases with increase in $N$, as evidenced by the scaling for $t$.
This is further elucidated in Fig.\ref{fig:gtNmax} where $G_{max}$ as a function of $N$ is shown. The maximum amplification is found to scale as $e^{12N} Re^{2}$ indicating an exponential increase with increasing value of the freestream shear. These results strongly indicate that a bypass transition scenario is likely to occur as the external freestream vorticity is increased.

\section{Concluding remarks}\label{sec:conc}

We studied the transient growth in a flat plate boundary layer under a uniform freestream shear, which is one the simplest problems involving external vorticity. The freestream vorticity introduces an additional second-order boundary layer effect. 
Though many aspects of the transient growth phenomenon were found to be similar with the Blasius flow, a larger optimum disturbance growth and a longer spanwise wavelength, in comparison with the Blasius flow is found.  
The maximum optimal amplification is found to increase with the freestream shear, with $G_{max}$ scaling exponentially with the freestream shear gradient. The present study is of interest in understanding the bypass transition mechanism in boundary layer flows subject to freestream vorticity, and should motivate further experimental research.

\section*{Acknowledgements}

We are grateful to Prof. Jyothirmoy Dey for introducing us to this interesting boundary layer flow, and the Department of Aerospace Engineering (Low Speed Wind Tunnel Lab) at the Indian Institute of Science, where the study began. 
S. S. G. acknowledges the support from the French National Research Agency (LABEX CEMPI, Grant No. ANR-11- LABX-0007) as well as the French Ministry of Higher Education and Research, Hauts de France council and European Regional Development Fund (ERDF) through the Contrat de Projets Etat-Region (CPER Photonics for Society P4S). A. C. M. thankfully acknowledges the financial support provided by IIT Kanpur. 

\nocite{*}
\bibliography{ICEAE_PhysFluids_Shear}

\begin{thebibliography}{49}%
\makeatletter
\providecommand \@ifxundefined [1]{%
 \@ifx{#1\undefined}
}%
\providecommand \@ifnum [1]{%
 \ifnum #1\expandafter \@firstoftwo
 \else \expandafter \@secondoftwo
 \fi
}%
\providecommand \@ifx [1]{%
 \ifx #1\expandafter \@firstoftwo
 \else \expandafter \@secondoftwo
 \fi
}%
\providecommand \natexlab [1]{#1}%
\providecommand \enquote  [1]{``#1''}%
\providecommand \bibnamefont  [1]{#1}%
\providecommand \bibfnamefont [1]{#1}%
\providecommand \citenamefont [1]{#1}%
\providecommand \href@noop [0]{\@secondoftwo}%
\providecommand \href [0]{\begingroup \@sanitize@url \@href}%
\providecommand \@href[1]{\@@startlink{#1}\@@href}%
\providecommand \@@href[1]{\endgroup#1\@@endlink}%
\providecommand \@sanitize@url [0]{\catcode `\\12\catcode `\$12\catcode
  `\&12\catcode `\#12\catcode `\^12\catcode `\_12\catcode `\%12\relax}%
\providecommand \@@startlink[1]{}%
\providecommand \@@endlink[0]{}%
\providecommand \url  [0]{\begingroup\@sanitize@url \@url }%
\providecommand \@url [1]{\endgroup\@href {#1}{\urlprefix }}%
\providecommand \urlprefix  [0]{URL }%
\providecommand \Eprint [0]{\href }%
\providecommand \doibase [0]{http://dx.doi.org/}%
\providecommand \selectlanguage [0]{\@gobble}%
\providecommand \bibinfo  [0]{\@secondoftwo}%
\providecommand \bibfield  [0]{\@secondoftwo}%
\providecommand \translation [1]{[#1]}%
\providecommand \BibitemOpen [0]{}%
\providecommand \bibitemStop [0]{}%
\providecommand \bibitemNoStop [0]{.\EOS\space}%
\providecommand \EOS [0]{\spacefactor3000\relax}%
\providecommand \BibitemShut  [1]{\csname bibitem#1\endcsname}%
\let\auto@bib@innerbib\@empty
\bibitem [{\citenamefont {Klebanoff}(1971)}]{klebanoff1971effect}%
  \BibitemOpen
  \bibfield  {author} {\bibinfo {author} {\bibfnamefont {P.}~\bibnamefont
  {Klebanoff}},\ }\bibfield  {title} {\enquote {\bibinfo {title} {Effect of
  free-stream turbulence on a laminar boundary layer},}\ }in\ \href@noop {}
  {\emph {\bibinfo {booktitle} {Bulletin of the American Physical Society}}},\
  Vol.~\bibinfo {volume} {16}\ (\bibinfo {year} {1971})\ p.\ \bibinfo {pages}
  {1323}\BibitemShut {NoStop}%
\bibitem [{\citenamefont {Kendall}(1998)}]{kendall1998experiments}%
  \BibitemOpen
  \bibfield  {author} {\bibinfo {author} {\bibfnamefont {J.}~\bibnamefont
  {Kendall}},\ }\bibfield  {title} {\enquote {\bibinfo {title} {Experiments on
  boundary-layer receptivity to freestream turbulence},}\ }in\ \href@noop {}
  {\emph {\bibinfo {booktitle} {36th AIAA Aerospace Sciences Meeting and
  Exhibit}}}\ (\bibinfo {year} {1998})\ p.\ \bibinfo {pages} {530}\BibitemShut
  {NoStop}%
\bibitem [{\citenamefont {Westin}\ \emph {et~al.}(1998)\citenamefont {Westin},
  \citenamefont {Bakchinov}, \citenamefont {Kozlov},\ and\ \citenamefont
  {Alfredsson}}]{westin1998experiments}%
  \BibitemOpen
  \bibfield  {author} {\bibinfo {author} {\bibfnamefont {K.}~\bibnamefont
  {Westin}}, \bibinfo {author} {\bibfnamefont {A.}~\bibnamefont {Bakchinov}},
  \bibinfo {author} {\bibfnamefont {V.}~\bibnamefont {Kozlov}}, \ and\ \bibinfo
  {author} {\bibfnamefont {P.}~\bibnamefont {Alfredsson}},\ }\bibfield  {title}
  {\enquote {\bibinfo {title} {Experiments on localized disturbances in a flat
  plate boundary layer. part 1. the receptivity and evolution of a localized
  free stream disturbance},}\ }\href@noop {} {\bibfield  {journal} {\bibinfo
  {journal} {European Journal of Mechanics-B/Fluids}\ }\textbf {\bibinfo
  {volume} {17}},\ \bibinfo {pages} {823--846} (\bibinfo {year}
  {1998})}\BibitemShut {NoStop}%
\bibitem [{\citenamefont {Saric}, \citenamefont {Reed},\ and\ \citenamefont
  {Kerschen}(2002)}]{saric2002boundary}%
  \BibitemOpen
  \bibfield  {author} {\bibinfo {author} {\bibfnamefont {W.~S.}\ \bibnamefont
  {Saric}}, \bibinfo {author} {\bibfnamefont {H.~L.}\ \bibnamefont {Reed}}, \
  and\ \bibinfo {author} {\bibfnamefont {E.~J.}\ \bibnamefont {Kerschen}},\
  }\bibfield  {title} {\enquote {\bibinfo {title} {Boundary-layer receptivity
  to freestream disturbances},}\ }\href@noop {} {\bibfield  {journal} {\bibinfo
   {journal} {Annual Review of Fluid Mechanics}\ }\textbf {\bibinfo {volume}
  {34}},\ \bibinfo {pages} {291--319} (\bibinfo {year} {2002})}\BibitemShut
  {NoStop}%
\bibitem [{\citenamefont {Schrader}\ \emph {et~al.}(2010)\citenamefont
  {Schrader}, \citenamefont {Brandt}, \citenamefont {Mavriplis},\ and\
  \citenamefont {Henningson}}]{schrader2010receptivity}%
  \BibitemOpen
  \bibfield  {author} {\bibinfo {author} {\bibfnamefont {L.-U.}\ \bibnamefont
  {Schrader}}, \bibinfo {author} {\bibfnamefont {L.}~\bibnamefont {Brandt}},
  \bibinfo {author} {\bibfnamefont {C.}~\bibnamefont {Mavriplis}}, \ and\
  \bibinfo {author} {\bibfnamefont {D.~S.}\ \bibnamefont {Henningson}},\
  }\bibfield  {title} {\enquote {\bibinfo {title} {Receptivity to free-stream
  vorticity of flow past a flat plate with elliptic leading edge},}\
  }\href@noop {} {\bibfield  {journal} {\bibinfo  {journal} {Journal of Fluid
  Mechanics}\ }\textbf {\bibinfo {volume} {653}},\ \bibinfo {pages} {245--271}
  (\bibinfo {year} {2010})}\BibitemShut {NoStop}%
\bibitem [{\citenamefont {Manu}, \citenamefont {Mathew},\ and\ \citenamefont
  {Dey}(2010)}]{Manu2010}%
  \BibitemOpen
  \bibfield  {author} {\bibinfo {author} {\bibfnamefont {K.~V.}\ \bibnamefont
  {Manu}}, \bibinfo {author} {\bibfnamefont {J.}~\bibnamefont {Mathew}}, \ and\
  \bibinfo {author} {\bibfnamefont {J.}~\bibnamefont {Dey}},\ }\bibfield
  {title} {\enquote {\bibinfo {title} {Evolution of isolated streamwise
  vortices in the late stages of boundary layer transition},}\ }\href@noop {}
  {\bibfield  {journal} {\bibinfo  {journal} {Experiments in Fluids}\ }\textbf
  {\bibinfo {volume} {48}},\ \bibinfo {pages} {431--440} (\bibinfo {year}
  {2010})}\BibitemShut {NoStop}%
\bibitem [{\citenamefont {Morkovin}(1969)}]{Mork69}%
  \BibitemOpen
  \bibfield  {author} {\bibinfo {author} {\bibfnamefont {M.~V.}\ \bibnamefont
  {Morkovin}},\ }\bibfield  {title} {\enquote {\bibinfo {title} {On the many
  faces of transition},}\ }in\ \href@noop {} {\emph {\bibinfo {booktitle}
  {Viscous Drag Reduction}}},\ \bibinfo {editor} {edited by\ \bibinfo {editor}
  {\bibfnamefont {C.~S.}\ \bibnamefont {Wells}}}\ (\bibinfo  {publisher}
  {Plenum},\ \bibinfo {year} {1969})\ pp.\ \bibinfo {pages} {1--31}\BibitemShut
  {NoStop}%
\bibitem [{\citenamefont {Jon{\'a}{\v{s}}}, \citenamefont {Mazur},\ and\
  \citenamefont {Uruba}(2000)}]{jonavs2000receptivity}%
  \BibitemOpen
  \bibfield  {author} {\bibinfo {author} {\bibfnamefont {P.}~\bibnamefont
  {Jon{\'a}{\v{s}}}}, \bibinfo {author} {\bibfnamefont {O.}~\bibnamefont
  {Mazur}}, \ and\ \bibinfo {author} {\bibfnamefont {V.}~\bibnamefont
  {Uruba}},\ }\bibfield  {title} {\enquote {\bibinfo {title} {On the
  receptivity of the by-pass transition to the length scale of the outer stream
  turbulence},}\ }\href@noop {} {\bibfield  {journal} {\bibinfo  {journal}
  {European Journal of Mechanics-B/Fluids}\ }\textbf {\bibinfo {volume} {19}},\
  \bibinfo {pages} {707--722} (\bibinfo {year} {2000})}\BibitemShut {NoStop}%
\bibitem [{\citenamefont {Henningson}(2006)}]{Henningson2006}%
  \BibitemOpen
  \bibfield  {author} {\bibinfo {author} {\bibfnamefont {D.}~\bibnamefont
  {Henningson}},\ }\bibfield  {title} {\enquote {\bibinfo {title} {Transient
  growth with application bypass transition to},}\ }in\ \href@noop {} {\emph
  {\bibinfo {booktitle} {IUTAM Symposium on Laminar-Turbulent Transition}}},\
  \bibinfo {editor} {edited by\ \bibinfo {editor} {\bibfnamefont
  {R.}~\bibnamefont {Govindarajan}}}\ (\bibinfo  {publisher} {Springer
  Netherlands},\ \bibinfo {address} {Dordrecht},\ \bibinfo {year} {2006})\ pp.\
  \bibinfo {pages} {15--24}\BibitemShut {NoStop}%
\bibitem [{\citenamefont {Van~Dyke}(1969)}]{van1969higher}%
  \BibitemOpen
  \bibfield  {author} {\bibinfo {author} {\bibfnamefont {M.}~\bibnamefont
  {Van~Dyke}},\ }\bibfield  {title} {\enquote {\bibinfo {title} {Higher-order
  boundary-layer theory},}\ }\href@noop {} {\bibfield  {journal} {\bibinfo
  {journal} {Annual Review of Fluid Mechanics}\ }\textbf {\bibinfo {volume}
  {1}},\ \bibinfo {pages} {265--292} (\bibinfo {year} {1969})}\BibitemShut
  {NoStop}%
\bibitem [{\citenamefont {Kovasznay}(1953)}]{Kova53}%
  \BibitemOpen
  \bibfield  {author} {\bibinfo {author} {\bibfnamefont {L.~S.~G.}\
  \bibnamefont {Kovasznay}},\ }\bibfield  {title} {\enquote {\bibinfo {title}
  {Turbulence in supersonic flow},}\ }\href@noop {} {\bibfield  {journal}
  {\bibinfo  {journal} {Journal of Aeronautical Sciences}\ }\textbf {\bibinfo
  {volume} {20}},\ \bibinfo {pages} {657--682} (\bibinfo {year}
  {1953})}\BibitemShut {NoStop}%
\bibitem [{\citenamefont {Crouch}(1994)}]{Crou94}%
  \BibitemOpen
  \bibfield  {author} {\bibinfo {author} {\bibfnamefont {J.~D.}\ \bibnamefont
  {Crouch}},\ }\bibfield  {title} {\enquote {\bibinfo {title} {Distributed
  excitation of {T}ollmien--{S}chlichting waves by vortical free-stream
  distribution},}\ }\href@noop {} {\bibfield  {journal} {\bibinfo  {journal}
  {Physics of Fluids}\ }\textbf {\bibinfo {volume} {6}},\ \bibinfo {pages}
  {217--223} (\bibinfo {year} {1994})}\BibitemShut {NoStop}%
\bibitem [{\citenamefont {Ovchinnikov}, \citenamefont {Piomelli},\ and\
  \citenamefont {Choudhari}(2006)}]{Ovch06}%
  \BibitemOpen
  \bibfield  {author} {\bibinfo {author} {\bibfnamefont {V.}~\bibnamefont
  {Ovchinnikov}}, \bibinfo {author} {\bibfnamefont {U.}~\bibnamefont
  {Piomelli}}, \ and\ \bibinfo {author} {\bibfnamefont {M.~M.}\ \bibnamefont
  {Choudhari}},\ }\bibfield  {title} {\enquote {\bibinfo {title} {Numerical
  simulations of boundary-layer transition induced by a cylinder wake},}\
  }\href@noop {} {\bibfield  {journal} {\bibinfo  {journal} {Journal of Fluid
  Mechanics}\ }\textbf {\bibinfo {volume} {547}},\ \bibinfo {pages} {413--441}
  (\bibinfo {year} {2006})}\BibitemShut {NoStop}%
\bibitem [{\citenamefont {Pan}\ \emph {et~al.}(2008)\citenamefont {Pan},
  \citenamefont {Wang}, \citenamefont {Zhang},\ and\ \citenamefont
  {H.}}]{Pan08}%
  \BibitemOpen
  \bibfield  {author} {\bibinfo {author} {\bibfnamefont {C.}~\bibnamefont
  {Pan}}, \bibinfo {author} {\bibfnamefont {J.~J.}\ \bibnamefont {Wang}},
  \bibinfo {author} {\bibfnamefont {P.~F.}\ \bibnamefont {Zhang}}, \ and\
  \bibinfo {author} {\bibfnamefont {F.~L.}\ \bibnamefont {H.}},\ }\bibfield
  {title} {\enquote {\bibinfo {title} {Coherent structures in bypass transition
  induced by a cylinder wake},}\ }\href@noop {} {\bibfield  {journal} {\bibinfo
   {journal} {Journal of Fluid Mechanics}\ }\textbf {\bibinfo {volume} {603}},\
  \bibinfo {pages} {367--389} (\bibinfo {year} {2008})}\BibitemShut {NoStop}%
\bibitem [{\citenamefont {Ferri}\ and\ \citenamefont
  {Libby}(1954)}]{ferri1954note}%
  \BibitemOpen
  \bibfield  {author} {\bibinfo {author} {\bibfnamefont {A.}~\bibnamefont
  {Ferri}}\ and\ \bibinfo {author} {\bibfnamefont {P.~A.}\ \bibnamefont
  {Libby}},\ }\bibfield  {title} {\enquote {\bibinfo {title} {Note on an
  interaction between the boundary layer and the inviscid flow},}\ }\href@noop
  {} {\bibfield  {journal} {\bibinfo  {journal} {Journal of the Aeronautical
  Sciences}\ }\textbf {\bibinfo {volume} {21}},\ \bibinfo {pages} {130--130}
  (\bibinfo {year} {1954})}\BibitemShut {NoStop}%
\bibitem [{\citenamefont {Li}(1956)}]{Li56}%
  \BibitemOpen
  \bibfield  {author} {\bibinfo {author} {\bibfnamefont {T.~V.}\ \bibnamefont
  {Li}},\ }\bibfield  {title} {\enquote {\bibinfo {title} {Effects of
  free-stream vorticity on the behavior of a viscous boundary layer},}\
  }\href@noop {} {\bibfield  {journal} {\bibinfo  {journal} {Journal of
  Aeronautical Sciences}\ }\textbf {\bibinfo {volume} {23}},\ \bibinfo {pages}
  {1128--1129} (\bibinfo {year} {1956})}\BibitemShut {NoStop}%
\bibitem [{\citenamefont {Glauert}(1957)}]{Glau57}%
  \BibitemOpen
  \bibfield  {author} {\bibinfo {author} {\bibfnamefont {M.~B.}\ \bibnamefont
  {Glauert}},\ }\bibfield  {title} {\enquote {\bibinfo {title} {The boundary
  layer in simple shear flow past a flat plate},}\ }\href@noop {} {\bibfield
  {journal} {\bibinfo  {journal} {Journal of Aeronautical Sciences}\ }\textbf
  {\bibinfo {volume} {24}},\ \bibinfo {pages} {848--849} (\bibinfo {year}
  {1957})}\BibitemShut {NoStop}%
\bibitem [{\citenamefont {Murray}(1961)}]{Murr61}%
  \BibitemOpen
  \bibfield  {author} {\bibinfo {author} {\bibfnamefont {J.~D.}\ \bibnamefont
  {Murray}},\ }\bibfield  {title} {\enquote {\bibinfo {title} {The boundary
  layer on a flat plate in a stream with uniform shear},}\ }\href@noop {}
  {\bibfield  {journal} {\bibinfo  {journal} {Journal of Fluid Mechanics}\
  }\textbf {\bibinfo {volume} {11}},\ \bibinfo {pages} {309--316} (\bibinfo
  {year} {1961})}\BibitemShut {NoStop}%
\bibitem [{\citenamefont {Devan}(1965)}]{devan1965approximate}%
  \BibitemOpen
  \bibfield  {author} {\bibinfo {author} {\bibfnamefont {L.}~\bibnamefont
  {Devan}},\ }\bibfield  {title} {\enquote {\bibinfo {title} {Approximate
  solution of the shear flow boundary layer on a flat plate},}\ }\href@noop {}
  {\bibfield  {journal} {\bibinfo  {journal} {Physics of Fluids}\ }\textbf
  {\bibinfo {volume} {8}},\ \bibinfo {pages} {2211--2215} (\bibinfo {year}
  {1965})}\BibitemShut {NoStop}%
\bibitem [{\citenamefont {Koch}, \citenamefont {Ludford},\ and\ \citenamefont
  {Seebass}(1971)}]{koch1971diffusion}%
  \BibitemOpen
  \bibfield  {author} {\bibinfo {author} {\bibfnamefont {W.}~\bibnamefont
  {Koch}}, \bibinfo {author} {\bibfnamefont {G.}~\bibnamefont {Ludford}}, \
  and\ \bibinfo {author} {\bibfnamefont {A.}~\bibnamefont {Seebass}},\
  }\bibfield  {title} {\enquote {\bibinfo {title} {Diffusion in shear flow past
  a semi-infinite flat plate},}\ }\href@noop {} {\bibfield  {journal} {\bibinfo
   {journal} {Acta Mechanica}\ }\textbf {\bibinfo {volume} {12}},\ \bibinfo
  {pages} {99--120} (\bibinfo {year} {1971})}\BibitemShut {NoStop}%
\bibitem [{\citenamefont {Dey}\ and\ \citenamefont {Nath}(1984)}]{Dey84}%
  \BibitemOpen
  \bibfield  {author} {\bibinfo {author} {\bibfnamefont {J.}~\bibnamefont
  {Dey}}\ and\ \bibinfo {author} {\bibfnamefont {G.}~\bibnamefont {Nath}},\
  }\bibfield  {title} {\enquote {\bibinfo {title} {A note on the incompressible
  flow past a semi-infinite flat plate in a stream with uniform shear},}\
  }\href@noop {} {\bibfield  {journal} {\bibinfo  {journal} {ASME Journal of
  Applied Mechanics}\ }\textbf {\bibinfo {volume} {51}},\ \bibinfo {pages}
  {210--211} (\bibinfo {year} {1984})}\BibitemShut {NoStop}%
\bibitem [{\citenamefont {Bera}\ and\ \citenamefont {Dey}(2005)}]{Bera05}%
  \BibitemOpen
  \bibfield  {author} {\bibinfo {author} {\bibfnamefont {N.}~\bibnamefont
  {Bera}}\ and\ \bibinfo {author} {\bibfnamefont {J.}~\bibnamefont {Dey}},\
  }\bibfield  {title} {\enquote {\bibinfo {title} {Linear instability of flow
  over a semi-infinite plate in a stream with uniform flow},}\ }\href@noop {}
  {\bibfield  {journal} {\bibinfo  {journal} {Acta Mechanica}\ }\textbf
  {\bibinfo {volume} {180}},\ \bibinfo {pages} {245--250} (\bibinfo {year}
  {2005})}\BibitemShut {NoStop}%
\bibitem [{\citenamefont {Legner}(2014)}]{legner2014optimal}%
  \BibitemOpen
  \bibfield  {author} {\bibinfo {author} {\bibfnamefont {H.~H.}\ \bibnamefont
  {Legner}},\ }\bibfield  {title} {\enquote {\bibinfo {title} {Optimal
  coordinates for higher-order boundary-layer theory},}\ }\href@noop {}
  {\bibfield  {journal} {\bibinfo  {journal} {Journal of Engineering
  Mathematics}\ }\textbf {\bibinfo {volume} {84}},\ \bibinfo {pages} {123--133}
  (\bibinfo {year} {2014})}\BibitemShut {NoStop}%
\bibitem [{\citenamefont {Balamurugan}\ and\ \citenamefont
  {Mandal}(2017{\natexlab{a}})}]{balamurugan2017experiments1}%
  \BibitemOpen
  \bibfield  {author} {\bibinfo {author} {\bibfnamefont {G.}~\bibnamefont
  {Balamurugan}}\ and\ \bibinfo {author} {\bibfnamefont {A.~C.}\ \bibnamefont
  {Mandal}},\ }\bibfield  {title} {\enquote {\bibinfo {title} {Experiments in
  bypass boundary layer transition under a stream with and without shear},}\
  }in\ \href@noop {} {\emph {\bibinfo {booktitle} {Journal of Physics:
  Conference Series}}},\ Vol.\ \bibinfo {volume} {822}\ (\bibinfo
  {organization} {IOP Publishing},\ \bibinfo {year} {2017})\ p.\ \bibinfo
  {pages} {012015}\BibitemShut {NoStop}%
\bibitem [{\citenamefont {Toomre}\ and\ \citenamefont
  {Rott}(1964)}]{toomre1964pressure}%
  \BibitemOpen
  \bibfield  {author} {\bibinfo {author} {\bibfnamefont {A.}~\bibnamefont
  {Toomre}}\ and\ \bibinfo {author} {\bibfnamefont {N.}~\bibnamefont {Rott}},\
  }\bibfield  {title} {\enquote {\bibinfo {title} {On the pressure induced by
  the boundary layer on a flat plate in shear flow},}\ }\href@noop {}
  {\bibfield  {journal} {\bibinfo  {journal} {Journal of Fluid Mechanics}\
  }\textbf {\bibinfo {volume} {19}},\ \bibinfo {pages} {1--10} (\bibinfo {year}
  {1964})}\BibitemShut {NoStop}%
\bibitem [{\citenamefont {Tumin}\ and\ \citenamefont
  {Reshotko}(2001)}]{tumin2001spatial}%
  \BibitemOpen
  \bibfield  {author} {\bibinfo {author} {\bibfnamefont {A.}~\bibnamefont
  {Tumin}}\ and\ \bibinfo {author} {\bibfnamefont {E.}~\bibnamefont
  {Reshotko}},\ }\bibfield  {title} {\enquote {\bibinfo {title} {Spatial theory
  of optimal disturbances in boundary layers},}\ }\href@noop {} {\bibfield
  {journal} {\bibinfo  {journal} {Physics of Fluids}\ }\textbf {\bibinfo
  {volume} {13}},\ \bibinfo {pages} {2097--2104} (\bibinfo {year}
  {2001})}\BibitemShut {NoStop}%
\bibitem [{\citenamefont {Fransson}\ \emph {et~al.}(2004)\citenamefont
  {Fransson}, \citenamefont {Brandt}, \citenamefont {Talamelli},\ and\
  \citenamefont {Cossu}}]{fransson2004experimental}%
  \BibitemOpen
  \bibfield  {author} {\bibinfo {author} {\bibfnamefont {J.~H.}\ \bibnamefont
  {Fransson}}, \bibinfo {author} {\bibfnamefont {L.}~\bibnamefont {Brandt}},
  \bibinfo {author} {\bibfnamefont {A.}~\bibnamefont {Talamelli}}, \ and\
  \bibinfo {author} {\bibfnamefont {C.}~\bibnamefont {Cossu}},\ }\bibfield
  {title} {\enquote {\bibinfo {title} {Experimental and theoretical
  investigation of the nonmodal growth of steady streaks in a flat plate
  boundary layer},}\ }\href@noop {} {\bibfield  {journal} {\bibinfo  {journal}
  {Physics of Fluids}\ }\textbf {\bibinfo {volume} {16}},\ \bibinfo {pages}
  {3627--3638} (\bibinfo {year} {2004})}\BibitemShut {NoStop}%
\bibitem [{\citenamefont {Mans}, \citenamefont {De~Lange},\ and\ \citenamefont
  {Van~Steenhoven}(2007)}]{mans2007sinuous}%
  \BibitemOpen
  \bibfield  {author} {\bibinfo {author} {\bibfnamefont {J.}~\bibnamefont
  {Mans}}, \bibinfo {author} {\bibfnamefont {H.}~\bibnamefont {De~Lange}}, \
  and\ \bibinfo {author} {\bibfnamefont {A.}~\bibnamefont {Van~Steenhoven}},\
  }\bibfield  {title} {\enquote {\bibinfo {title} {Sinuous breakdown in a flat
  plate boundary layer exposed to free-stream turbulence},}\ }\href@noop {}
  {\bibfield  {journal} {\bibinfo  {journal} {Physics of Fluids}\ }\textbf
  {\bibinfo {volume} {19}},\ \bibinfo {pages} {088101} (\bibinfo {year}
  {2007})}\BibitemShut {NoStop}%
\bibitem [{\citenamefont {Duriez}, \citenamefont {Aider},\ and\ \citenamefont
  {Wesfreid}(2009)}]{duriez2009self}%
  \BibitemOpen
  \bibfield  {author} {\bibinfo {author} {\bibfnamefont {T.}~\bibnamefont
  {Duriez}}, \bibinfo {author} {\bibfnamefont {J.-L.}\ \bibnamefont {Aider}}, \
  and\ \bibinfo {author} {\bibfnamefont {J.~E.}\ \bibnamefont {Wesfreid}},\
  }\bibfield  {title} {\enquote {\bibinfo {title} {Self-sustaining process
  through streak generation in a flat-plate boundary layer},}\ }\href@noop {}
  {\bibfield  {journal} {\bibinfo  {journal} {Physical Review Letters}\
  }\textbf {\bibinfo {volume} {103}},\ \bibinfo {pages} {144502} (\bibinfo
  {year} {2009})}\BibitemShut {NoStop}%
\bibitem [{\citenamefont {Balamurugan}\ and\ \citenamefont
  {Mandal}(2017{\natexlab{b}})}]{balamurugan2017experiments2}%
  \BibitemOpen
  \bibfield  {author} {\bibinfo {author} {\bibfnamefont {G.}~\bibnamefont
  {Balamurugan}}\ and\ \bibinfo {author} {\bibfnamefont {A.~C.}\ \bibnamefont
  {Mandal}},\ }\bibfield  {title} {\enquote {\bibinfo {title} {Experiments on
  localized secondary instability in bypass boundary layer transition},}\
  }\href@noop {} {\bibfield  {journal} {\bibinfo  {journal} {Journal of Fluid
  Mechanics}\ }\textbf {\bibinfo {volume} {817}},\ \bibinfo {pages} {217--263}
  (\bibinfo {year} {2017}{\natexlab{b}})}\BibitemShut {NoStop}%
\bibitem [{\citenamefont {Vavaliaris}, \citenamefont {Beneitez},\ and\
  \citenamefont {Henningson}(2020)}]{vavaliaris2020optimal}%
  \BibitemOpen
  \bibfield  {author} {\bibinfo {author} {\bibfnamefont {C.}~\bibnamefont
  {Vavaliaris}}, \bibinfo {author} {\bibfnamefont {M.}~\bibnamefont
  {Beneitez}}, \ and\ \bibinfo {author} {\bibfnamefont {D.~S.}\ \bibnamefont
  {Henningson}},\ }\bibfield  {title} {\enquote {\bibinfo {title} {Optimal
  perturbations and transition energy thresholds in boundary layer shear
  flows},}\ }\href@noop {} {\bibfield  {journal} {\bibinfo  {journal} {Physical
  Review Fluids}\ }\textbf {\bibinfo {volume} {5}},\ \bibinfo {pages} {062401}
  (\bibinfo {year} {2020})}\BibitemShut {NoStop}%
\bibitem [{\citenamefont {Rigas}, \citenamefont {Sipp},\ and\ \citenamefont
  {Colonius}(2021)}]{rigas2021nonlinear}%
  \BibitemOpen
  \bibfield  {author} {\bibinfo {author} {\bibfnamefont {G.}~\bibnamefont
  {Rigas}}, \bibinfo {author} {\bibfnamefont {D.}~\bibnamefont {Sipp}}, \ and\
  \bibinfo {author} {\bibfnamefont {T.}~\bibnamefont {Colonius}},\ }\bibfield
  {title} {\enquote {\bibinfo {title} {Nonlinear input/output analysis:
  application to boundary layer transition},}\ }\href@noop {} {\bibfield
  {journal} {\bibinfo  {journal} {Journal of Fluid Mechanics}\ }\textbf
  {\bibinfo {volume} {911}} (\bibinfo {year} {2021})}\BibitemShut {NoStop}%
\bibitem [{\citenamefont {Drazin}\ and\ \citenamefont
  {Reid}(2004)}]{drazin2004hydrodynamic}%
  \BibitemOpen
  \bibfield  {author} {\bibinfo {author} {\bibfnamefont {P.~G.}\ \bibnamefont
  {Drazin}}\ and\ \bibinfo {author} {\bibfnamefont {W.~H.}\ \bibnamefont
  {Reid}},\ }\href@noop {} {\emph {\bibinfo {title} {Hydrodynamic stability}}}\
  (\bibinfo  {publisher} {Cambridge university press},\ \bibinfo {year}
  {2004})\BibitemShut {NoStop}%
\bibitem [{\citenamefont {Schmid}(2007)}]{schmid2007nonmodal}%
  \BibitemOpen
  \bibfield  {author} {\bibinfo {author} {\bibfnamefont {P.~J.}\ \bibnamefont
  {Schmid}},\ }\bibfield  {title} {\enquote {\bibinfo {title} {Nonmodal
  stability theory},}\ }\href@noop {} {\bibfield  {journal} {\bibinfo
  {journal} {Annual Review of Fluid Mechanics}\ }\textbf {\bibinfo {volume}
  {39}},\ \bibinfo {pages} {129--162} (\bibinfo {year} {2007})}\BibitemShut
  {NoStop}%
\bibitem [{\citenamefont {Farrell}(1988{\natexlab{a}})}]{Farr88}%
  \BibitemOpen
  \bibfield  {author} {\bibinfo {author} {\bibfnamefont {B.~F.}\ \bibnamefont
  {Farrell}},\ }\bibfield  {title} {\enquote {\bibinfo {title} {Optimal
  excitation of perturbations in viscous shear flow},}\ }\href@noop {}
  {\bibfield  {journal} {\bibinfo  {journal} {Physics of Fluids}\ }\textbf
  {\bibinfo {volume} {31}},\ \bibinfo {pages} {2093--2102} (\bibinfo {year}
  {1988}{\natexlab{a}})}\BibitemShut {NoStop}%
\bibitem [{\citenamefont {Farrell}(1988{\natexlab{b}})}]{farrell1988optimal}%
  \BibitemOpen
  \bibfield  {author} {\bibinfo {author} {\bibfnamefont {B.~F.}\ \bibnamefont
  {Farrell}},\ }\bibfield  {title} {\enquote {\bibinfo {title} {Optimal
  excitation of perturbations in viscous shear flow},}\ }\href@noop {}
  {\bibfield  {journal} {\bibinfo  {journal} {Physics of Fluids}\ }\textbf
  {\bibinfo {volume} {31}},\ \bibinfo {pages} {2093--2102} (\bibinfo {year}
  {1988}{\natexlab{b}})}\BibitemShut {NoStop}%
\bibitem [{\citenamefont {Reddy}\ and\ \citenamefont
  {Henningson}(1993)}]{Redd93}%
  \BibitemOpen
  \bibfield  {author} {\bibinfo {author} {\bibfnamefont {S.~C.}\ \bibnamefont
  {Reddy}}\ and\ \bibinfo {author} {\bibfnamefont {D.~S.}\ \bibnamefont
  {Henningson}},\ }\bibfield  {title} {\enquote {\bibinfo {title} {Energy
  growth in viscous channel flows},}\ }\href@noop {} {\bibfield  {journal}
  {\bibinfo  {journal} {Journal of Fluid Mechanics}\ }\textbf {\bibinfo
  {volume} {252}},\ \bibinfo {pages} {209--238} (\bibinfo {year}
  {1993})}\BibitemShut {NoStop}%
\bibitem [{\citenamefont {Butler}\ and\ \citenamefont
  {Farrell}(1992)}]{Butl92}%
  \BibitemOpen
  \bibfield  {author} {\bibinfo {author} {\bibfnamefont {K.~M.}\ \bibnamefont
  {Butler}}\ and\ \bibinfo {author} {\bibfnamefont {B.~F.}\ \bibnamefont
  {Farrell}},\ }\bibfield  {title} {\enquote {\bibinfo {title}
  {Three-dimensional optimal perturbations in viscous shear flow},}\
  }\href@noop {} {\bibfield  {journal} {\bibinfo  {journal} {Physics of
  Fluids}\ }\textbf {\bibinfo {volume} {4}},\ \bibinfo {pages} {1637--1650}
  (\bibinfo {year} {1992})}\BibitemShut {NoStop}%
\bibitem [{\citenamefont {Schmid}(2000)}]{Schm00}%
  \BibitemOpen
  \bibfield  {author} {\bibinfo {author} {\bibfnamefont {P.~J.}\ \bibnamefont
  {Schmid}},\ }\bibfield  {title} {\enquote {\bibinfo {title} {Linear stability
  theory and bypass transition in shear flow},}\ }\href@noop {} {\bibfield
  {journal} {\bibinfo  {journal} {Physics of Plasmas}\ }\textbf {\bibinfo
  {volume} {7}},\ \bibinfo {pages} {1788--1794} (\bibinfo {year}
  {2000})}\BibitemShut {NoStop}%
\bibitem [{\citenamefont {Ellingsen}\ and\ \citenamefont
  {Palm}(1975)}]{Elli75}%
  \BibitemOpen
  \bibfield  {author} {\bibinfo {author} {\bibfnamefont {T.}~\bibnamefont
  {Ellingsen}}\ and\ \bibinfo {author} {\bibfnamefont {E.}~\bibnamefont
  {Palm}},\ }\bibfield  {title} {\enquote {\bibinfo {title} {Stability of
  linear flow},}\ }\href@noop {} {\bibfield  {journal} {\bibinfo  {journal}
  {Physics of Fluids}\ }\textbf {\bibinfo {volume} {18}},\ \bibinfo {pages}
  {487--488} (\bibinfo {year} {1975})}\BibitemShut {NoStop}%
\bibitem [{\citenamefont {Landahl}(1980)}]{Land80}%
  \BibitemOpen
  \bibfield  {author} {\bibinfo {author} {\bibfnamefont {M.~T.}\ \bibnamefont
  {Landahl}},\ }\bibfield  {title} {\enquote {\bibinfo {title} {A note on an
  algebraic instability of inviscid parallel shear flows},}\ }\href@noop {}
  {\bibfield  {journal} {\bibinfo  {journal} {Journal of Fluid Mechanics}\
  }\textbf {\bibinfo {volume} {98}},\ \bibinfo {pages} {243--251} (\bibinfo
  {year} {1980})}\BibitemShut {NoStop}%
\bibitem [{\citenamefont {Hultgren}\ and\ \citenamefont
  {Gustavsson}(1981)}]{Hult81}%
  \BibitemOpen
  \bibfield  {author} {\bibinfo {author} {\bibfnamefont {L.~S.}\ \bibnamefont
  {Hultgren}}\ and\ \bibinfo {author} {\bibfnamefont {L.~H.}\ \bibnamefont
  {Gustavsson}},\ }\bibfield  {title} {\enquote {\bibinfo {title} {Algebraic
  growth of disturbances in a laminar boundary layer},}\ }\href@noop {}
  {\bibfield  {journal} {\bibinfo  {journal} {Physics of Fluids}\ }\textbf
  {\bibinfo {volume} {24}},\ \bibinfo {pages} {1000--1004} (\bibinfo {year}
  {1981})}\BibitemShut {NoStop}%
\bibitem [{\citenamefont {Andersson}, \citenamefont {Berggren},\ and\
  \citenamefont {Henningson}(1999)}]{Ande99}%
  \BibitemOpen
  \bibfield  {author} {\bibinfo {author} {\bibfnamefont {P.}~\bibnamefont
  {Andersson}}, \bibinfo {author} {\bibfnamefont {M.}~\bibnamefont {Berggren}},
  \ and\ \bibinfo {author} {\bibfnamefont {D.~S.}\ \bibnamefont {Henningson}},\
  }\bibfield  {title} {\enquote {\bibinfo {title} {Optimal disturbances and
  bypass transition in boundary layers},}\ }\href@noop {} {\bibfield  {journal}
  {\bibinfo  {journal} {Physics of Fluids}\ }\textbf {\bibinfo {volume} {11}},\
  \bibinfo {pages} {134--150} (\bibinfo {year} {1999})}\BibitemShut {NoStop}%
\bibitem [{\citenamefont {Gustavsson}(1991)}]{Gust91}%
  \BibitemOpen
  \bibfield  {author} {\bibinfo {author} {\bibfnamefont {L.~H.}\ \bibnamefont
  {Gustavsson}},\ }\bibfield  {title} {\enquote {\bibinfo {title} {Energy
  growth of three-dimensional disturbances in plane {P}oiseuille flow},}\
  }\href@noop {} {\bibfield  {journal} {\bibinfo  {journal} {Journal of Fluid
  Mechanics}\ }\textbf {\bibinfo {volume} {224}},\ \bibinfo {pages} {241--260}
  (\bibinfo {year} {1991})}\BibitemShut {NoStop}%
\bibitem [{\citenamefont {Corbett}\ and\ \citenamefont
  {Bottaro}(2000)}]{Corb00}%
  \BibitemOpen
  \bibfield  {author} {\bibinfo {author} {\bibfnamefont {P.}~\bibnamefont
  {Corbett}}\ and\ \bibinfo {author} {\bibfnamefont {A.}~\bibnamefont
  {Bottaro}},\ }\bibfield  {title} {\enquote {\bibinfo {title} {Optimal
  perturbations for boundary layers subject to stream-wise pressure
  gradient},}\ }\href@noop {} {\bibfield  {journal} {\bibinfo  {journal}
  {Physics of Fluids}\ }\textbf {\bibinfo {volume} {12}},\ \bibinfo {pages}
  {120--130} (\bibinfo {year} {2000})}\BibitemShut {NoStop}%
\bibitem [{\citenamefont {Luchini}(2000)}]{luchini2000reynolds}%
  \BibitemOpen
  \bibfield  {author} {\bibinfo {author} {\bibfnamefont {P.}~\bibnamefont
  {Luchini}},\ }\bibfield  {title} {\enquote {\bibinfo {title}
  {Reynolds-number-independent instability of the boundary layer over a flat
  surface: optimal perturbations},}\ }\href@noop {} {\bibfield  {journal}
  {\bibinfo  {journal} {Journal of Fluid Mechanics}\ }\textbf {\bibinfo
  {volume} {404}},\ \bibinfo {pages} {289--309} (\bibinfo {year}
  {2000})}\BibitemShut {NoStop}%
\bibitem [{\citenamefont {H{\oe}pffner}, \citenamefont {Brandt},\ and\
  \citenamefont {Henningson}(2005)}]{hoepffner2005transient}%
  \BibitemOpen
  \bibfield  {author} {\bibinfo {author} {\bibfnamefont {J.}~\bibnamefont
  {H{\oe}pffner}}, \bibinfo {author} {\bibfnamefont {L.}~\bibnamefont
  {Brandt}}, \ and\ \bibinfo {author} {\bibfnamefont {D.~S.}\ \bibnamefont
  {Henningson}},\ }\bibfield  {title} {\enquote {\bibinfo {title} {Transient
  growth on boundary layer streaks},}\ }\href@noop {} {\bibfield  {journal}
  {\bibinfo  {journal} {Journal of Fluid Mechanics}\ }\textbf {\bibinfo
  {volume} {537}},\ \bibinfo {pages} {91--100} (\bibinfo {year}
  {2005})}\BibitemShut {NoStop}%
\bibitem [{\citenamefont {Zuccher}, \citenamefont {Bottaro},\ and\
  \citenamefont {Luchini}(2006)}]{zuccher2006algebraic}%
  \BibitemOpen
  \bibfield  {author} {\bibinfo {author} {\bibfnamefont {S.}~\bibnamefont
  {Zuccher}}, \bibinfo {author} {\bibfnamefont {A.}~\bibnamefont {Bottaro}}, \
  and\ \bibinfo {author} {\bibfnamefont {P.}~\bibnamefont {Luchini}},\
  }\bibfield  {title} {\enquote {\bibinfo {title} {Algebraic growth in a
  {B}lasius boundary layer: Nonlinear optimal disturbances},}\ }\href@noop {}
  {\bibfield  {journal} {\bibinfo  {journal} {European Journal of
  Mechanics-B/Fluids}\ }\textbf {\bibinfo {volume} {25}},\ \bibinfo {pages}
  {1--17} (\bibinfo {year} {2006})}\BibitemShut {NoStop}%
\bibitem [{\citenamefont {Schmid}\ and\ \citenamefont
  {Henningson}(2001)}]{Schm01}%
  \BibitemOpen
  \bibfield  {author} {\bibinfo {author} {\bibfnamefont {P.~J.}\ \bibnamefont
  {Schmid}}\ and\ \bibinfo {author} {\bibfnamefont {D.~S.}\ \bibnamefont
  {Henningson}},\ }in\ \href@noop {} {\emph {\bibinfo {booktitle} {{Stability
  and Transition in Shear Flows}}}}\ (\bibinfo  {publisher} {Springer-Verlag},\
  \bibinfo {address} {New York},\ \bibinfo {year} {2001})\BibitemShut {NoStop}%
\end{thebibliography}%

\end{document}